\documentclass[aps,prd,twocolumn,nofootinbib,superscriptaddress]{revtex4-2}

\usepackage{graphicx}
\usepackage{xcolor}
\usepackage{tikz}
\usepackage{hyperref}
\usepackage{amsmath,amsfonts,amssymb}
\usepackage{orcidlink}

\definecolor{cyan_dp}{HTML}{007272}
\definecolor{rose}{HTML}{a84847}
\definecolor{green}{HTML}{6A8023}
\definecolor{orange}{HTML}{FF6600}
\definecolor{red}{HTML}{FF2511}
\definecolor{blue}{HTML}{055C9D}
\definecolor{orchid}{HTML}{9c3386}
\definecolor{ref_blue}{HTML}{00478F}

\hypersetup{
    colorlinks,
    citecolor=ref_blue,
    filecolor=black,
    linkcolor=orchid,
    urlcolor=cyan_dp
}

\newcommand\nmesh{\texttt{Nmesh}}
\newcommand\ba{\begin{eqnarray}}
\newcommand\ea{\end{eqnarray}}
\newcommand\be{\begin{equation}}
\newcommand\ee{\end{equation}}
\newcommand\p{{\partial}}
\newcommand{\ts}{\textsubscript}
\newcommand{\ds}{\displaystyle}

\begin{document}

\title{Neutron star evolution by combining discontinuous Galerkin and
finite volume methods}

\author{Ananya Adhikari\orcidlink{0000-0002-3890-3577}}
\affiliation{Department of Physics, Florida Atlantic University,
                    Boca Raton, FL 33431, USA}
\affiliation{CENTRA, Departamento de F\'isica, Instituto Superior
  T\'ecnico IST, Universidade de Lisboa UL, Avenida Rovisco Pais 1,
  1049 Lisboa, Portugal}

\author{Wolfgang Tichy\orcidlink{0000-0002-8707-754X}}
\affiliation{Department of Physics, Florida Atlantic University,
                    Boca Raton, FL 33431, USA}

\author{Liwei Ji\orcidlink{0000-0001-9008-0267}}
\affiliation{Center for Computational Relativity and Gravitation,
                    \& School of Mathematical Sciences,
Rochester Institute of Technology, Rochester, New York 14623, USA}

\author{Amit Poudel\orcidlink{0000-0001-5752-8218}}
\affiliation{Department of Physics, Florida Atlantic University,
                    Boca Raton, FL 33431, USA}

\date{\today}

\begin{abstract}

We present here a new hybrid scheme that combines a
discontinuous Galerkin (DG) method with compact finite volume (FV) and
finite difference (FD) methods.
The computational mesh is divided into smaller
elements that touch but do not overlap. Like a pure DG method, our new
hybrid scheme requires information exchange only at the surface of
neighboring elements. This avoids the need for ghost zones that are usually
many points deep in traditional FV implementations. Furthermore, unlike
traditional FV implementations, that require information exchange between
each element and its 26 surrounding neighbors on noncuboid meshes, our new
hybrid method exchanges information only between each element and its six
nearest neighbors. Through this reduction in communication, we aim to retain
the high scalability of DG when using large supercomputers.
In addition, the information exchange between adjacent elements is much
simpler than in a traditional FV implementation, because we always have
grid points at the interface, so that only surface interpolation is required.
As a result it is much easier to implement adaptive mesh refinement.
The goal is to
use DG in elements with smooth matter fields and to fall back onto the more
robust FV/FD method in elements that contain nonsmooth shocks or star
surfaces. For this we devise trouble criteria to decide whether an element
should be evolved with DG or FV/FD.
We use the \nmesh~program to implement and test the new scheme. We
successfully evolve various single neutron star cases. These include the
challenging cases of a neutron star initially in an unstable equilibrium
migrating to a stable configuration and a boosted neutron star. These cases
are simulated for the first time here in full 3D with general relativistic
hydrodynamics using DG methods. We also describe additional numerical
methods, such as the limiters and the atmosphere treatment we need for our
simulations.

\end{abstract}

\maketitle


\section{Introduction}

The first gravitational wave (GW) signal observation in 2015 heralded the
hundredth year commemoration of Einstein's formulation of the theory of general
relativity~\cite{Einstein15a} and the associated equations of
gravity~\cite{Einstein15b}.
This historic observation of a binary black hole (BBH)
merger~\cite{Abbott:2016blz,LIGOScientific:2016emj}
by the Laser Interferometer Gravitational-Wave
Observatory (LIGO) and the Virgo interferometer
opened a new avenue of
studying the universe through the GW spectrum. While
other BBH merger observations~\cite{Abbott:2016nmj,Abbott:2017vtc} followed,
the first binary neutron star (BNS) merger was
observed~\cite{TheLIGOScientific:2017qsa} only in 2017.
This event, designated GW170817 by
LIGO collaboration, was also observed across the electromagnetic (EM)
spectrum~\cite{GBM:2017lvd,LIGOScientific:2017zic,Coulter:2017wya}. This
observation of an EM counterpart for the GW signal
launched the current epoch of multimessenger astrophysics. The possibility
of observing BNS mergers across multiple spectra make them especially
interesting as GW sources.
EM counterparts from BBH mergers can only
be observed if the black holes are embedded
in a dense matter environment that
gets charged and heated during the inspiral and
merger~\cite{Schnittman:2010wy,Kelly:2017xck,Graham:2020gwr,Chen:2023xrm},
and hence are much rarer.
BNS mergers and black hole-neutron star
mergers~\cite{LIGOScientific:2021qlt} are thus much
more promising and lucrative candidates for multimessenger astrophysics.
Furthermore, studying the neutron star (NS) related events also allows
us to probe the related properties and phenomena
such as equation of state (EoS) of matter at supranuclear
densities~\cite{Radice:2017lry,Huth:2021bsp,Golomb:2024mmt},
production of heavy elements via rapid neutron capture (r-process)
nucleosynthesis~\cite{Metzger:2019zeh,Curtis:2021guz,Ricigliano:2024lwf},
and value of the Hubble
constant~\cite{Dietrich:2020efo,Kunert:2024xqb}.

The sensitivity of the GW detectors has continuously
increased~\cite{LIGOScientific:2018mvr,
LIGOScientific:2020ibl,LIGOScientific:2021djp,Capote:2024rmo}. In addition,
the LIGO-Virgo-KAGRA~\cite{KAGRA:2018plz,Aso:2013eba,Somiya:2011np}
detectors are projected to further increase in
sensitivity~\cite{Aasi:2013wya}.
More importantly, the sensitivity of
the upcoming next generation~\cite{Bailes:2021tot} of GW detectors, namely
Cosmic Explorer~\cite{Reitze:2019iox,Evans:2016mbw},
the DECi-hertz Interferometer Gravitational-wave Observatory
(DECIGO)~\cite{Kawamura:2020pcg},
Einstein Telescope~\cite{Punturo:2010zz},
LIGO Voyager~\cite{Adhikari:2019zpy},
the Laser Interferometer Space Antenna (LISA)~\cite{LISA:2017pwj},
NEMO~\cite{Ackley:2020atn}, and
TianQin~\cite{TianQin:2015yph} is expected to be yet much higher.
We thus can look forward to GW observations with even higher
signal-to-noise ratio.

GW signals observed are analyzed using
template banks~\cite{Sakon:2022ibh,Kumar:2013gwa} which in turn
use results from databases of numerical simulations of
compact binary coalescences (CBCs), such as the CoRe~\cite{Gonzalez:2022mgo}, SXS~\cite{Boyle:2019kee},
and RIT~\cite{Healy:2019jyf} databases.
While older generation programs such as
\texttt{BAM}~\cite{Bruegmann:2006ulg,Thierfelder:2011yi,Dietrich:2015iva},
\texttt{Einstein Toolkit}~\cite{Loeffler2012,EinsteinToolkit:2020_11},
\texttt{SACRA-MPI}~\cite{Kiuchi:2019kzt},
NRPy+~\cite{Ruchlin:2017com}, and
\texttt{SpEC}~\cite{SpECWebsite,Boyle:2019kee} provide reliable
simulation results, they cannot easily meet the demands for accuracy and
speed for the upcoming GW detectors~\cite{Wang:2024iyj,Puecher:2022oiz,
Williams:2022vct,Breschi:2021xrx,Purrer:2019jcp,Hall2019a}.
In some cases this happens because a program may lack important features
such as well-parallelized and efficient adaptive mesh refinement (AMR).
Yet in most cases,
it is due to large overheads in the numerical
methods used in older generation programs.
The usual practice is to decompose the computational mesh into
smaller domains or elements and to distribute these elements
among available compute cores.
However, since the physical system spans the entire mesh,
the elements still need to communicate information between them.
The large overhead comes from the need to exchange information
many grid points deep, between nearest neighbor and sometimes even next to
nearest neighbor elements, when one uses traditional finite volume (FV) or
finite differencing (FD) methods.
Additionally, for FV and FD schemes, achieving higher-order error
convergence is challenging, necessitating high-resolution simulations,
which in turn increase the cost again. Hence, fully utilizing the
computing capability of modern high-performance computers (HPCs),
that contain hundreds of thousands or even million of compute cores,
becomes difficult for these traditional methods.

These challenges have motivated the creation of a new generation
of numerical relativity (NR) programs, some of which we list here.
Discontinuous Galerkin methods are used in
\texttt{bamps}~\cite{Bugner:2015gqa,Hilditch:2015aba},
\texttt{ExaHyPE}~\cite{Koppel:2017kiz,Fambri:2018udk}, and
\texttt{SpECTRE}~\cite{Kidder:2016hev,Deppe:2021bhi}.
Others like \texttt{CarpetX}~\cite{CarpetX_web},
\texttt{AsterX}~\cite{Kalinani:2024rbk},
\texttt{GRaM-X}~\cite{Shankar:2022ful},
\texttt{GR-Athena++}~\cite{Daszuta:2021ecf},
\texttt{AthenaK}~\cite{Zhu:2024utz},
\texttt{Parthenon}~\cite{Grete2022a},
\texttt{GRChombo}~\cite{Clough:2015sqa,Andrade:2021rbd},
and \texttt{Dendro-GR}~\cite{Fernando:2018mov},
explore leveraging graphics processing units
while using FV or FD methods. \texttt{Dendro-GR} additionally
uses wavelet-based adaptive mesh refinement.
\verb|SPHINCS_BSSN|~\cite{Rosswog:2020kwm}
uses smooth particle hydrodynamics for matter evolution and
FD for spacetime evolution.

\nmesh~\cite{Tichy:2022hpa} aims to be such a next-generation NR program.
It uses a
discontinuous Galerkin (DG) method to achieve its goal of high scalability
and accuracy when running on modern HPCs.
Introduced first by Reed and Hill~\cite{Reed1973a} in 1973, the
DG method took its rigorous form in a series of breakthrough
articles by Cockburn and Shu~\cite{Cockburn:1991a,Cockburn:1989a,
Cockburn:1989b,Cockburn:1990a,Cockburn1998a}.
Although historically employed in engineering and
astrophysics, recent years have seen a growing interest in DG
in the NR community~\cite{Radice:2011qr,Bugner:2015gqa,Teukolsky:2015ega,
Hilditch:2015aba,
Kidder:2016hev,Dumbser:2017okk,Koppel:2017kiz,Fambri:2018udk,
Cao:2018myc,Deppe:2021ada,
Deppe:2021bhi}.
DG offers two advantages: high-order convergence
when the fields are smooth and less communication overhead
between neighboring elements distributed among different compute cores.
The first comes from its spectral nature and the second comes from the
nature of its boundary conditions between neighboring elements.

However, simulation of neutron stars with full general relativistic
hydrodynamics (GRHD) contain
shocks and nonsmooth features in the matter fields, in addition to
smooth regions. While DG can
handle shocks and nonsmooth features between mesh element boundaries,
it fails to adequately resolve them when these
lie in the interior of an element. Such cases result in
spurious Gibbs oscillations~\cite{Wilbraham1848a,Gibbs1898a,
Gibbs1899a,Hewitt1979} that drastically reduce accuracy.
Additionally, DG is still subject to Godunov's
theorem~\cite{Godunov1959a} when dealing with discontinuities
and thus has low-order convergence when they
are present.
Hence, supplemental methods for
handling these nonsmooth features become necessary. While many nonlinear
limiters~\cite{Shu:1987a,Cockburn:1989a,Cockburn1998a,Schaal:2015ila,
Krivodonova2007a,Moe:2015a} have been proposed for this purpose,
they fall short~\cite{Deppe:2021ada,Deppe:2021bhi,Deppe:2024ckt}
compared to the more robust traditional FV methods, with e.g.~weighted
essentially nonoscillatory (WENO)~\cite{Liu1994a,Jiang1996a}
reconstruction, that are used for high-resolution shock capturing (HRSC).
This has motivated a different approach to the problem by using hybrid
schemes that take advantage of DG's higher accuracy at lower computational
cost in regions where the matter fields are smooth, while falling back
to the more robust FV+WENO formalism when handling nonsmooth fields.
Whether an element is evolved with DG or FV is determined by some
sort of ``trouble'' detection criteria.
The first implementation of this kind, to the best of our knowledge is
presented in~\cite{Costa2007a}. Subsequent examples of similar
implementations can be found in~\cite{Huerta2012a,Sonntag2014a,
Dumbser2014a,Boscheri2017a,Zanotti:2015mia,Fambri:2018udk,
NunezdelaRosa2018a,Deppe:2021ada,Deppe:2024ckt}. A significant achievement
in this approach, that has since been adopted by subsequent works,
was the \emph{a posteriori} approach in~\cite{Dumbser2014a}.
As in, trouble detection criteria are applied after a time step
has been done. Then, if an element is troubled, and has been evolved with
DG, the element is projected on to the FV or FD domain and the time step is
redone. This is a significant difference from the limiters,
which essentially tried to remove spurious
Gibbs oscillations~\cite{Wilbraham1848a,Gibbs1898a,Gibbs1899a,Hewitt1979}
that inevitably arise when shocks are present. These, however,
are better handled by FV methods with HRSC.
Since the step is not redone to start from the
physically sound values, the limiters approach inevitably retains some
unphysciality that can grow over time, which is in contrast to the
\emph{a posteriori} approach.

In this paper, we present a new FV/FD method that easily combines with the
DG method. Our new scheme needs the same amount of communication as the DG
method, and hence does not suffer from the communication overhead problem of
traditional FV schemes described above.
Part of our new FV/FD method is a new WENO reconstruction scheme,
which we call ``WENOZ\ts{2},''
that uses a higher order WENOZ~\cite{Borges:2008a} method in the interior
of an element and then drops the reconstruction order
first to WENO3~\cite{Liu1994a} and then to even lower order
as the element boundary is approached.
We also devise certain trouble criteria
(similar as in~\cite{Deppe:2021ada})
to detect elements with smooth matter fields, and use an
\emph{a posteriori} hybrid method that uses DG in smooth regions, but
switches elements with nonsmooth matter to use our new FV/FD method.

We have implemented the new hybrid DG+FV/FD scheme in our \nmesh~program. To
test it, we simulate four single initial neutron star setups: a static star
in a stable Tolman-Oppenheimer-Volkoff (TOV)~\cite{Tolman39,Oppenheimer39b}
configuration, a perturbed star, a highly dense star initially in an
unstable equilibrium configuration that migrates towards a stable
configuration, and a boosted neutron star. To the best of our knowledge,
these are the first successful fully 3D simulations of an unstable star
migrating to a stable state, and of a boosted neutron star using DG methods. The DG
implementation remains the same as presented in our first
paper~\cite{Tichy:2022hpa}. We build on the single neutron star simulations
with a static metric (Cowling approximation)~\cite{Cowling1941} presented
there and simulate all the stars in this article with a dynamic metric. We
find that our new hybrid scheme works well in all cases tested.

In Sec.~\ref{grhd}, we briefly restate the equations of GRHD, for evolving
both the metric and matter variables. Sec.~\ref{methods} describes our
new FV/FD and WENOZ\ts{2} methods along with other necessary numerical
approaches we use. We present the results from preliminary test cases
of a scalar field and a blast wave evolution and then the evolution
of the four initial neutron star setups in Sec.~\ref{results}.
Last, we state our conclusions in Sec.~\ref{conclusions}.

We note that the elements that make up the computational mesh of
\nmesh~were called leaf nodes or just nodes in most parts
of~\cite{Tichy:2022hpa}, since these elements are the leaf nodes in an
octree, that is used to track mesh refinement. However, in the present paper
we have decided to use their more conventional name, and we will call them
elements.

We use dimensionless units where $G=c=M_{\odot}=1$. To convert to Système international
(SI) units, a dimensionless length has to be
multiplied by $L_0 = 1476.6250$~m,
a time by $T_0 = 4.9254909\times10^{-6}$~s,
a mass by $M_{\odot}=1.9884099\times10^{30}$~kg,
and a mass density by $6.1758285\times10^{20}$~kg/m$^3$.
All quantities in this article are expressed
in these units unless explicitly stated otherwise.
For tensor notations, indices from Latin alphabet,
such as $i$, run from $1$ to $3$
and denote spatial indices, while Greek indices,
such as $\mu$, run from $0$
to $3$ and denote spacetime indices.


\section{Equations of general relativistic hydrodynamics}
\label{grhd}

The evolution equations for both the metric system 
and the hydrodynamic variables remain the same
as that outlined in Secs.~4.2 and 4.3 of~\cite{Tichy:2022hpa}.
However, we restate them here for the sake of completeness.

\subsection{Metric evolution}

We use the standard 3+1 decomposition of the 4-metric~\cite{Arnowitt62}
\be
\label{4metric}
ds^2 = g_{\mu\nu}dx^{\mu}dx^{\nu}
= -\alpha^2 dt^2 + \gamma_{ij}(dx^i + \beta^i dt)(dx^j + \beta^j dt) ,
\ee
where $\alpha$, $\beta^i$, and $\gamma_{ij}$ are lapse, shift, and
3-metric. The unit normal to a hypersurface of constant coordinate
time $t$ is then given by $n_{\alpha} = (-\alpha,0,0,0)$.

To evolve the gravitational fields, we use the generalized harmonic (GH)
system~\cite{Lindblom:2005qh,Hilditch:2015aba} given by
\ba
\label{eq:GHG_g}
&\partial_t g_{\alpha\beta}&
-(1+\gamma_1)\beta^k\partial_kg_{\alpha\beta} =
-\alpha\Pi_{\alpha\beta}-\gamma_1\beta^k\Phi_{k\alpha\beta}, \\
\label{eq:GHG_Pi}
&\partial_t\Pi_{\alpha\beta}&
-\beta^k\partial_k\Pi_{\alpha\beta}
+\alpha\gamma^{ki}\partial_k\Phi_{i\alpha\beta}
-\gamma_1\gamma_2\beta^k\partial_kg_{\alpha\beta} = \nonumber \\
&&2\alpha g^{\mu \nu}
\left(\gamma^{ij}\Phi_{i\mu\alpha}\Phi_{j\nu\beta}-\Pi_{\mu\alpha}\Pi_{\nu\beta}
-g^{\sigma \rho}\Gamma_{\alpha \mu \sigma}\Gamma_{\beta \nu \rho}\right)
 \nonumber\\
&&-2\alpha\nabla_{(\alpha}H_{\beta)}-\frac{1}{2}\alpha n^\mu
n^\nu\Pi_{\mu \nu}\Pi_{\alpha\beta}
-\alpha n^\mu\Pi_{\mu i}\gamma^{ij}\Phi_{j\alpha\beta} \nonumber\\
&&+\alpha\gamma_0\left[2\delta^\mu{}_{(\alpha}n_{\beta)}
-g_{\alpha\beta}n^\mu\right]
\left(H_\mu+\Gamma_\mu\right)
 \nonumber\\
&&-\gamma_1\gamma_2\beta^k\Phi_{k\alpha\beta} -16\pi\alpha
\left(
  T_{\alpha\beta}-\frac{1}{2}g_{\alpha\beta}g^{\mu \nu}T_{\mu \nu}
\right), 
\ea
\ba
&\partial_t\Phi_{i\alpha\beta}&
-\beta^k\partial_k\Phi_{i\alpha\beta}+\alpha\partial_i\Pi_{\alpha\beta}
-\alpha\gamma_2\partial_ig_{\alpha\beta} = \nonumber \\
&&\frac{1}{2}\alpha n^\mu n^\nu\Phi_{i\mu \nu}\Pi_{\alpha\beta}
+\alpha\gamma^{jk}n^\mu
\Phi_{ij\mu}\Phi_{k\alpha\beta} \nonumber \\
&&-\alpha\gamma_2\Phi_{i\alpha\beta}. \label{eq:GHG_Phi}
\ea
Here $g_{\alpha\beta}$ is the spacetime metric
and $\Gamma_\alpha=g^{\mu \nu}\Gamma_{\alpha \mu \nu}$
is the contracted Christoffel symbol.
The equations are written in terms of the extra variables
$\Pi_{\alpha\beta}:=-n^\mu\p_\mu g_{\alpha\beta}$
and $\Phi_{i\alpha\beta}:=\p_ig_{\alpha\beta}$,
that have been introduced to make the original second order GH system
first order in both time and space. Gauge conditions in the GH system are
specified by prescribing the gauge source function $H_\alpha$.
We state our gauge choices for each of our simulation setups explicitly
in Sec.~\ref{results}.
The lapse $\alpha$, shift $\beta^i$ and spatial metric $\gamma_{ij}$ come
from the $3+1$ decomposition in Eq.~(\ref{4metric}).

The GH evolution equations also contain extra terms
for constraint damping that are multiplied with
the factors $\gamma_{0}$, $\gamma_{1}$, and $\gamma_{2}$.
For some of the runs, we set these factors to
the constant values of $\gamma_{0}=0.1$, $\gamma_{1}=-1$, and $\gamma_{2}=1$.
Yet in most other cases, we set them to~\cite{Deppe:2024ckt}
\begin{equation}
\label{3gaussians}
\gamma_{i}(\vec{x}) = C_{(i)}
 + \sum_{\alpha = 1}^3 A^{(i)}_{(\alpha)}
   \exp\left(- \ds \frac{ \left(\vec{x} - \vec{x}_{(\alpha)} \right)^2 }
			{(w^{(i)}_{(\alpha)})^2}  \right)
,
\end{equation}
so that they become functions of the position vector $\vec{x}$. Here
$\vec{x}_{(1)}$ and $\vec{x}_{(2)}$ are the centers of stars 1 and 2,
while $\vec{x}_{(3)}$ is the center of mass (CoM).
The values of the parameters
$C_{(i)}$, $A^{(i)}_{(\alpha)}$ and $w^{(i)}_{(\alpha)}$ 
are given in Eqs.~(\ref{Par3Gf})-(\ref{Par3Gl}).

\subsection{Evolution of hydrodynamic fields}
\label{hydro}

To treat neutron star matter, we use the Valencia formulation~\cite{Banyuls97}.
Matter is thus described as a perfect fluid,
where the stress-energy tensor is given by
\be
\label{eq:Tmunu}
  T^{\alpha\beta} = \left(\rho + P \right) u^\alpha u^\beta
	+ P g^{\alpha \beta} .
\ee
Here $\rho$ is the energy density, $P$ is the pressure,
and $u^\mu$ is the four-velocity.
The total energy density is written as
\be
\rho=\rho_0(1+\epsilon) ,
\ee
where $\rho_0$ is the rest-mass energy density,
and $\epsilon$ is the specific internal energy.
We express the four-velocity $u^\mu$ in terms of the three-velocity
given by
\be
v^{\mu} = u^{\mu}/W - n^{\mu}
\ee
and also introduce the Lorentz factor
\be
W = -n_{\mu} u^{\mu} = \alpha u^0.
\ee
Together $(\rho_0, \epsilon, Wv^i, P)$ are known as the primitive variables.

The evolution equations for the matter are typically brought
into conservative form
\be
\label{Cons1}
\partial_t u +  \partial_i f^i = s ,
\ee
where $u$ denotes the conserved matter variables with corresponding fluxes
and sources $f^i$ and $s$.

In the Valencia formulation the conserved variables are
\ba
\label{D-def}
D    &=& \rho_0 W, \\
\label{tau-def}
\tau &=& \rho_0 h W^2 - P - \rho_0 W, \\
\label{Si-def}
S_i  &=& \rho_0 h W^2 v_i.
\ea
Here $D$ is rest-mass density, $\tau$ the internal energy density, and $S_i$ the
momentum density as seen by Eulerian observers. The last two equations
also contain the specific enthalpy given by
\be
h = 1+\epsilon + P/\rho_0 .
\ee
The conserved variables that correspond to $u$ in Eq.~(\ref{Cons1}) are
then
\be
u = \sqrt{\gamma}
\label{GRHDcons}
\left(
\begin{array}{ccc}
D    \\
\tau \\
S_l  \\
\end{array}
\right).
\ee
\begin{widetext}
They satisfy Eq.~(\ref{Cons1}), with the flux vectors and sources given by
\be
\label{GRHDfluxes}
f^i = \sqrt{\gamma}
\left(
\begin{array}{ccc}
(\alpha v^i - \beta^i) D				\\
(\alpha v^i - \beta^i) \tau + \alpha P v^i		\\
(\alpha v^i - \beta^i) S_l + \alpha P \delta^{i}_l	\\
\end{array}
\right)
\ee
and
\be
\label{GRHDsources}
\frac{s}{\sqrt{\gamma}} =
\left(
\begin{array}{ccc}
0 \\
T^{00} (\beta^i \beta^j K_{ij} - \beta^i \partial_i\alpha) +
	T^{0i} (2\beta^j K_{ij} - \partial_i\alpha)
	+ T^{ij} K_{ij}	\\
T^{00}(\frac{\beta^i\beta^j}{2}\partial_l\gamma_{ij} - \alpha\partial_l\alpha)+
	T^{0i} \beta^j \partial_l\gamma_{ij}
	+ T^0_i \partial_l\beta^i
	+ \frac{T^{ij}}{2}\partial_l\gamma_{ij}	\\
\end{array}
\right).
\ee
\end{widetext}

The components of the stress-energy tensor appearing here can be expressed
in terms of the primitive variables as
\ba
T^{00} &=& (W^2 h\rho_0 - P)/\alpha^2, \\
T^{0i} &=& W h\rho_0 u^i/\alpha   + P\beta^i/\alpha^2, \\
T^{ij} &=& h \rho_0 u^i u^j + P (\gamma^{ij} - \beta^i\beta^j/\alpha^2), \\
T^0_i  &=& h \rho_0 W^2 v_i/\alpha,
\ea
where
\be
u^i = W v^i - W\frac{\beta^i}{\alpha}.
\ee

To close the evolution system, we have to specify an EoS
for the fluid, i.e.~an equation of the form
\be
\label{eos_general}
P = P(\rho_0, \epsilon)
\ee
that allows us to obtain the pressure for a given rest-mass energy density
and the specific internal energy, as well as the sound speed squared
$c^2_\mathrm{s}$.
If $c^2_\mathrm{s}<0$ or $c^2_\mathrm{s}>1$, we set it to zero.
We also set it to zero if $\rho_0=0$ or $h=0$. 
For the initial data of all the simulations
discussed in Sec.~\ref{results} we use a polytropic EoS
\be
\label{inidataeos}
P = \ds \kappa \rho_0^{1+\frac{1}{n}} ~\mathrm{and} ~\epsilon
= n \kappa \rho_0^{\frac{1}{n}},
\ee
with $\kappa = 100$ and $n = 1$. However, the systems are then evolved with
an ideal or ``gamma-law'' EoS
\be
\label{eos}
\ds P = \ds \frac{1}{n} \rho_0 \epsilon ,
\ee
that is of the form in Eq.~(\ref{eos_general}), which allows matter
to heat up.


\section{Numerical methods}
\label{methods}

We use the same DG implementation that we use in~\cite{Tichy:2022hpa}.
However, we find that this DG method is not enough for
simulating cases of boosted stars with relatively higher
velocities or unstable stars that rapidly expand or contract. 
For matter fields with fast moving nonsmooth features or even shocks,
the much more robust FV method with e.g.~WENO~\cite{Liu1994a} is necessary.
Regions containing nonsmooth features in the matter fields are classified as
``troubled''. In these troubled regions we will use an FV method for the
matter fields and an FD method for the gravitational fields.
We will refer to this combination as the ``FV/FD'' method.
However, we still want to use DG for all fields in nontroubled regions
to leverage its higher accuracy and lower computational cost.
Hence, we opt for a hybrid DG+FV/FD method.
This has been explored in
other works before~\cite{Huerta2012a,Sonntag2014a,Dumbser2014a,
Koppel:2017kiz,Deppe:2021ada}, but the specifics of combining the two 
schemes differ between those works and ours. Our FV/FD is more compatible 
with our DG implementation and also our goal of high scalability.


\subsection{The DG method}

As explained in~\cite{Tichy:2022hpa}, a DG method can be used for any system
of evolution equations that is first order in space, no matter if it is in
conservative form or not. The only difference will be the precise form
of the discretized evolution equations.

For a set of conservative equations of the form~(\ref{Cons1})
with a flux $f^i$, the DG discretization takes the form
\ba
\label{Integ4}
  \partial_t u_{q_1 q_2 q_3} 
 &+& \sum_{r=0}^N \Bigg(
    \frac{\partial x^{\bar{1}}}{\partial x^i}
    D_{q_1 r}^{\bar{1}} f^i_{r q_2 q_3}  \nonumber\\
 &+& \frac{\partial x^{\bar{2}}}{\partial x^i}
    D_{q_2 r}^{\bar{2}} f^i_{q_1 r q_3}  
    + \frac{\partial x^{\bar{3}}}{\partial x^i} 
    D_{q_3 r}^{\bar{3}} f^i_{q_1 q_2 r}
  \Bigg)  \nonumber\\
 = s_{q_1 q_2 q_3}
 &-&\frac{\sqrt{\bar{\gamma}^{\bar{1}\bar{1}}_{q_1 q_2 q_3}}}{w_{q_1}}
  F_{q_1 q_2 q_3} (\delta_{q_1 0} + \delta_{q_1 N})  \nonumber\\
  &-&\frac{\sqrt{\bar{\gamma}^{\bar{2}\bar{2}}_{q_1 q_2 q_3}}}{w_{q_2}}
  F_{q_1 q_2 q_3} (\delta_{q_2 0} + \delta_{q_2 N})  \nonumber\\
  &-&\frac{\sqrt{\bar{\gamma}^{\bar{3}\bar{3}}_{q_1 q_2 q_3}}}{w_{q_3}}
  F_{q_1 q_2 q_3} (\delta_{q_3 0} + \delta_{q_3 N}) ,
\ea
where
\be
\label{def_F}
F := (f^i n_i)^* - f^i n_i ,
\ee
and $(f^i n_i)^*$ denotes the numerical flux, computed from the
flux in the local element and the adjacent element. To arrive at
Eq.~(\ref{Integ4}) we have performed Gau{\ss}ian quadrature using
$N+1$ Gau\ss-Lobatto grid points in each of the three directions.
The grid point coordinates in each local element are denoted by
$x^{\bar{i}}_{q_1 q_2 q_3}$ where $q_i \in [0,N]$ is the grid point index
in each direction $i$. Here $u_{q_1 q_2 q_3}$ and $f^i_{q_1 q_2 q_3}$
denote $u$ and $f^i$ at grid point $x^{\bar{i}}_{q_1 q_2 q_3}$,
$D_{qr}^{\bar{i}}$ is the differentiation matrix in the
$x^{\bar{i}}$-direction,
$\frac{\partial x^{\bar{i}}}{\partial x^i}$ represents the coordinate
transformation between local element coordinates $x^{\bar{i}}$ and global
Cartesian-like coordinates $x^i$,
and $w_q$ is the Gau\ss~quadrature weight (see~\cite{Tichy:2022hpa}).

On the other hand, for a set of nonconservative equations of the form
\be
\label{NonCons}
\partial_t u + A^i(u) \partial_i u = s .
\ee
the DG formulation results in
\ba
\label{IntegA}
    \partial_t u_{q_1 q_2 q_3}
  &+& \sum_{r=0}^N A^i_{q_1 q_2 q_3} \Bigg(
    \frac{\partial x^{\bar{1}}}{\partial x^i}
    D_{q_1 r}^{\bar{1}} u_{r q_2 q_3} \nonumber\\
  &+& \frac{\partial x^{\bar{2}}}{\partial x^i}
    D_{q_2 r}^{\bar{2}} u_{q_1 r q_3}
  + \frac{\partial x^{\bar{3}}}{\partial x^i}
    D_{q_3 r}^{\bar{3}} u_{q_1 q_2 r}
  \Bigg)  \nonumber\\
= s_{q_1 q_2 q_3}
 &-&\frac{\sqrt{\bar{\gamma}^{\bar{1}\bar{1}}_{q_1 q_2 q_3}}}{w_{q_1}}
  G_{q_1 q_2 q_3} (\delta_{q_1 0} + \delta_{q_1 N}) \nonumber\\
 &-&\frac{\sqrt{\bar{\gamma}^{\bar{2}\bar{2}}_{q_1 q_2 q_3}}}{w_{q_2}}
  G_{q_1 q_2 q_3} (\delta_{q_2 0} + \delta_{q_2 N})  \nonumber\\
 &-&\frac{\sqrt{\bar{\gamma}^{\bar{3}\bar{3}}_{q_1 q_2 q_3}}}{w_{q_3}}
  G_{q_1 q_2 q_3} (\delta_{q_3 0} + \delta_{q_3 N}) ,
\ea
where $G$ is defined by
\be
\label{def_G}
G := (n_i A^i u)^* - n_i A^i u .
\ee

When we evolve neutron stars, the evolution equations for the matter fields
in Eq.~(\ref{GRHDcons}) obey a conservative equation of the form
(\ref{Cons1}), while the evolution equations for gravity are given by
Eqs.~(\ref{eq:GHG_g}), (\ref{eq:GHG_Pi}) and (\ref{eq:GHG_Phi}), which are
not in conservative form. Thus, we use the DG formulation in
Eq.(\ref{Integ4}) for the matter fields and that in
Eq.(\ref{IntegA}) for the gravitational fields.
Note that such DG schemes are highly accurate for smooth $u$, but
will not work as well for shocks.


\subsection{Modified finite volume / finite difference method}
\label{modFV}

In order to deal with shocks or other nonsmooth behavior such as star
surfaces it is often better to use finite volume or finite
difference methods. Here we explain how we can modify Eqs.~(\ref{Integ4})
and (\ref{IntegA}) for such cases.
First we will change our mesh and use $N+1$ equally spaced grid points
with grid spacing $h=2/N$.
\begin{figure}
  \includegraphics[width=\linewidth]{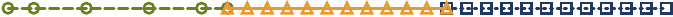}
  \caption{\label{gridpoints}
  Grid points in 1D. On the left is one DG element, with two
  FV/FD elements to its right that have uniform grid spacing.}
\end{figure}
As in Fig.~\ref{gridpoints}, we still place grid points at the boundary.

For nonconservative systems we can then use the same discretization as in
Eq.~(\ref{IntegA}). However, we modify the
differentiation matrix $D_{qr}^{\bar{i}}$ so that it corresponds to an
FD method using only three points. At or close to the
element boundary we use one-sided stencils, so that all derivatives can be
computed locally, without information from neighboring elements. This results
in a band diagonal matrix
\footnote{On its own this matrix does not possess the summation by
parts property discussed in e.g.~\cite{Calabrese:2003vx}. However,
Eq.~(\ref{IntegA}) has extra terms on its right-hand side, which act as
penalty terms, that help to stabilize the method.}
given by
\be
\sum_{r=0}^N D_{qr}^{\bar{i}} u_r = \left\{
{\renewcommand{\arraystretch}{2.5}%
\begin{array}{ll}
\ds \frac{1}{2h}(-3u_{0} + 4u_{1} - u_{2})		& \mbox{if $q=0$}, \\
\ds \frac{1}{2h}(-u_{q-1} + u_{q+1})		& \mbox{if $0<q<N$}, \\
\ds \frac{1}{2h}(u_{N-2} - 4u_{N-1} + 3u_{N})	& \mbox{if $q=N$} .
\end{array}
}
\right.
\ee
The quadrature weights are set to
\be
w_q = \left\{
\begin{array}{ll}
h/2  & \mbox{if $q=0$ or $q=N$},  \\
h    & \mbox{otherwise} ,
\end{array}
\right.
\ee
which corresponds to the trapezoidal integration rule.
The final modification is that we add Kreiss-Oliger dissipation terms of
the form
\be
\label{KO-diss}
-\frac{\sigma}{16h}
\sum_{r=0}^N  ( d_{q_1 r}^{\bar{1}} u_{r q_2 q_3}
               +d_{q_2 r}^{\bar{2}} u_{q_1 r q_3}
               +d_{q_3 r}^{\bar{3}} u_{q_1 q_2 r} )
\ee
to the right-hand side (RHS) of Eq.~(\ref{IntegA}). Here $\sigma$ is a dissipation factor
that we usually choose to be of order $10^{-1}$. The matrix
\begin{widetext}
\be
\sum_{r=0}^N d_{qr}^{\bar{i}} u_r = \left\{
\begin{array}{ll}
  \label{diss-matrix}
u_{q-2} - 4u_{q-1} + 6u_{q} - 4u_{q+1} + u_{q+2}& \mbox{if $1<q<N-1$} \\
0						& \mbox{otherwise}
\end{array}
\right.
\ee
\end{widetext}
is chosen such that dissipation occurs only in the interior of the element.
We use this scheme for the evolution equations of gravity given by
Eqs.~(\ref{eq:GHG_g}), (\ref{eq:GHG_Pi}) and (\ref{eq:GHG_Phi}), when
we use equally spaced grids.

For conservative systems of the form~(\ref{Cons1})
we use an FV scheme given by
\ba
\label{FV-scheme}
\partial_t u_{q_1 q_2 q_3}
+\frac{1}{J_{q_1 q_2 q_3}}\Bigg\{&&
\left[\frac{d(J \sqrt{\bar{\gamma}^{\bar{1}\bar{1}}} n^{\bar{1}}_i f^i)}
             {dx^{\bar{1}}}\right]_{q_1 q_2 q_3} \nonumber\\
 &+&\left[\frac{d(J \sqrt{\bar{\gamma}^{\bar{2}\bar{2}}} n^{\bar{2}}_i f^i)}
             {dx^{\bar{2}}}\right]_{q_1 q_2 q_3} \nonumber\\
 &+&\left[\frac{d(J \sqrt{\bar{\gamma}^{\bar{3}\bar{3}}} n^{\bar{3}}_i f^i)}
             {dx^{\bar{3}}}\right]_{q_1 q_2 q_3}
\Bigg\} \nonumber \\
= s_{q_1 q_2 q_3} &&.
\ea
For interior points ($0<q_1<N$) the first square bracket term is defined as
\ba
&& \left[\frac{d(J \sqrt{\bar{\gamma}^{\bar{1}\bar{1}}} n^{\bar{1}}_i f^i)}
           {dx^{\bar{1}}}\right]_{q_1} \nonumber\\
&:=&
     \frac{
         \left[ J\sqrt{\bar{\gamma}^{\bar{1}\bar{1}}} (n^{\bar{1}}_i  f^{i})^*
             \right]_{q_1+\frac{1}{2}}
        +\left[ J\sqrt{\bar{\gamma}^{\bar{1}\bar{1}}} (n^{\bar{1}}_i  f^{i})^*
             \right]_{q_1-\frac{1}{2}}
          }{h} ,\nonumber \\
          && \label{FV-term1}
\ea
with the trivial dependence on $q_2, q_3$ suppressed.
Here $n^{\bar{i}}_i$ denotes the outward pointing normal of
the element boundary $x^{\bar{i}}=\pm 1$, and
\be
J = \left|\det\left(
     \frac{\partial x^{\bar{i}}}{\partial x^i}
    \right)\right|^{-1} .
\ee
Notice that the flux derivatives are written here in terms
of values halfway between grid points, e.g. points
$x^{\bar{1}}_{q_1+\frac{1}{2}\ q_2\ q_3}$ and
$x^{\bar{1}}_{q_1-\frac{1}{2}\ q_2\ q_3}$. This means we have to obtain
the flux $f^i$ at locations between grid points by interpolation.
As one can see from Eq.~(\ref{FV-scheme}), this interpolation can be
done separately in each direction. Let us focus on the term
in Eq.~(\ref{FV-term1}).
The numerical flux $(n^{\bar{1}}_i  f^{i})^*$ in this expression is obtained
by interpolating to the midpoints $ \ds x^{\bar{1}}_{q_1\pm\frac{1}{2}}$.
For points far enough from the boundary we can use standard schemes
such as WENO3~\cite{Liu1994a} or WENOZ~\cite{Borges:2008a}.
Let us now consider $q_1=1$, i.e. one point in from
the left boundary. In this case
$(n^{\bar{1}}_i  f^{i})^*_{1+\frac{1}{2}}$ can be obtained by simple
WENO3 interpolation using data at the points with indices 0, 1, and 2.
On the other hand,
$(n^{\bar{1}}_i  f^{i})^*_{1-\frac{1}{2}}$ is obtained by linear
interpolation using data at the points with indices 0 and 1.
This means that, near the boundary, our scheme is only second order
accurate (error of $O(h^2)$) because of the use of linear interpolation.
Finally, we consider $q_1=0$, i.e. the point at the left boundary.
In this case we first compute
\ba
\label{nearbound}
&& \left[
\frac{d(J \sqrt{\bar{\gamma}^{\bar{1}\bar{1}}} n^{\bar{1}}_i f^i)}
     {dx^{\bar{1}}}\right]_{\frac{1}{4}} \nonumber\\
&:=&
     \frac{
         \left[ J\sqrt{\bar{\gamma}^{\bar{1}\bar{1}}} (n^{\bar{1}}_i  f^{i})^*
             \right]_{\frac{1}{2}}
        +\left[ J\sqrt{\bar{\gamma}^{\bar{1}\bar{1}}} (n^{\bar{1}}_i  f^{i})^*
             \right]_{0}
          }{h/2} .
\ea
Here $(n^{\bar{1}}_i  f^{i})^*_{\frac{1}{2}}$ is again obtained by linear
interpolation using data at the points with indices 0 and 1.
For $(n^{\bar{1}}_i  f^{i})^*_{0}$ the situation is different, as it is
located directly at the boundary where we do have grid points. Here
we use the same scheme as in the DG method and compute the numerical flux
from the value in this element and the adjacent element.

\begin{figure*}
  \centering
	\begin{tikzpicture}
    \foreach \x in {1.25,3.75,6.25,8.75,11.25,13.75}
    {
      \draw[thick, dashed] (\x,-1) -- (\x,1); 

      \pgfmathsetmacro{\result}{\x == 1.25}
      \ifdim\result pt=1pt
        \node[orange] at (\x+0.5,-0.9) {$\mathrm{W_-}$};
      \else
        \pgfmathsetmacro{\result}{\x == 13.75}
        \ifdim\result pt=1pt
          \node[orange] at (\x-0.4,-0.9) {$\mathrm{W_+}$};
        \else
          \pgfmathsetmacro{\result}{\x == 3.75}
          \ifdim\result pt=1pt
            \node[orange] at (\x-0.4,-0.9) {$\mathrm{W_+}$};
            \node[green] at (\x+0.5,-0.9) {$\mathrm{W_-}$};
          \else
            \pgfmathsetmacro{\result}{\x == 11.25}
            \ifdim\result pt=1pt
              \node[green] at (\x-0.4,-0.9) {$\mathrm{W_+}$};
              \node[orange] at (\x+0.5,-0.9) {$\mathrm{W_-}$};
            \else
              \node[green] at (\x+0.5,-0.9) {$\mathrm{W_-}$};
              \node[green] at (\x-0.4,-0.9) {$\mathrm{W_+}$};
            \fi
          \fi
        \fi
      \fi
    }
    \draw[thick] (0,-1) -- (0,1);
    \draw[thick] (15,-1) -- (15,1);
    \draw[->, thick, orange] (1.95,-1.5)  -- (1.75,-1.1);
    \draw[->, thick, orange] (2.95,-1.5)  -- (3.15,-1.1);
    \draw[->, thick, green] (6.65,-1.8)  -- (4.25,-1.1);
    \draw[->, thick, green] (6.65,-1.6)  -- (5.85,-1.1);
    \draw[->, thick, green] (6.95,-1.5)  -- (6.75,-1.1);
    \draw[->, thick, green] (7.95,-1.5)  -- (8.15,-1.1);
    \draw[->, thick, green] (8.35,-1.6)  -- (9.15,-1.1);
    \draw[->, thick, green] (8.35,-1.8)  -- (10.75,-1.1);
    \draw[->, thick, orange] (11.95,-1.5)  -- (11.75,-1.1);
    \draw[->, thick, orange] (12.95,-1.5)  -- (13.15,-1.1);
    \node[orange, below] at  (2.5,-1.5) {WENO3};
    \node[green, below] at  (7.5,-1.5) {WENOZ};
    \node[orange, below] at  (12.5,-1.5) {WENO3};
    \node[rose] at (1,-0.85) {L};
    \node[rose] at (14,-0.85) {L};
    \draw[->, thick, rose] (1,-2.1)  -- (1,-1.1);
    \draw[->, thick, rose] (14,-2.15)  -- (14,-1.1);
    \node[rose , below right] at (0.6,-2.05) {Linear};
    \node[rose , below left] at (14.4,-2.05) {Linear};
    \node[gray] at (0.12,-1.3) {$\mathrm{A}_w$};
    \node[gray] at (15.1,-1.3) {$\mathrm{A}_w$};
    \draw[->, thick, gray] (0.0,-2.7)  -- (0.0,-1.5);
    \draw[->, thick, gray] (15,-2.75)  -- (15,-1.5);
    \node[gray , below right] at (-0.3,-2.65) {Weighted Average};
    \node[gray , below left] at (15.3,-2.65) {Weighted Average};
    \foreach \x/\i in {0.0/0,2.5/1,5.0/2,7.5/3,10.0/4,12.5/5,15.0/6}
    {
      \draw[line width=1.5pt,blue] (\x,0) circle (3pt);
      \node[blue] at (\x,1.25) {$\i$};
    }
    \draw[red, line width = 1.25] (0.525,-0.1) -- (0.725,0.1);
    \draw[red, line width = 1.25] (0.525,0.1) -- (0.725,-0.1);
    \draw[red, line width = 1.25] (14.225,-0.1) -- (14.425,0.1);
    \draw[red, line width = 1.25] (14.225,0.1) -- (14.425,-0.1);
    \draw[orange, line width = 1.25] (-0.015,-0.25) -- (5.015,-0.25);
    \draw[orange, line width = 1.25] (5,-0.265) -- (5,0);
    \draw[orange, line width = 1.25] (2.5,-0.25) -- (2.5,0);
    \draw[orange, line width = 1.25] (0,-0.265) -- (0,0);
    \draw[->, thick, orange] (3.2,-0.6)  -- (3.2,-0.25);
    \draw[green, line width = 1.25] (4.985,0.25) -- (15.015,0.25);
    \draw[green, line width = 1.25] (5,0.265) -- (5,0);
    \draw[green, line width = 1.25] (7.5,0.25) -- (7.5,0);
    \draw[green, line width = 1.25] (10,0.25) -- (10,0);
    \draw[green, line width = 1.25] (12.5,0.25) -- (12.5,0);
    \draw[green, line width = 1.25] (15,0.265) -- (15,0);
    \draw[->, thick, green] (9.15,-0.6)  -- (9.15,0.25);
  \end{tikzpicture}
  \caption{\label{FVWENO}
  1D grid point arrangement for FV with WENO
  reconstruction. The numbering at the top shows the grid point
  $q_1$ = 0, 1, \dots . The dashed lines indicate the midpoints between
  grid points. The region between two consecutive midpoints is called a
  cell. For each cell we need to reconstruct fluxes at its left and right
  end. In the interior of the grid (in cells $q_1$ = 2, \dots, $N-2$),
  when we have five consecutive grid points available for interpolation,
  we use WENOZ reconstruction.
  In cells $q_1 = 1$ and $q_1 = (N-1)$ we drop the order of interpolation
  and use WENO3 instead. One stencil for each of WENOZ and WENO3 has been
  shown in the figure to illustrate their actual usage of the grid points.
  Even closer to the boundary,
  e.g.~in the cell between grid point 0 and midpoint $\frac{1}{2}$
  (marked by \textcolor{red}{$\times$}), we perform linear interpolation
  using fields at $q_1 = 0$ and $q_1 = 1$ to obtain fluxes at
  midpoint $\frac{1}{2}$. The same is done at midpoint $N-\frac{1}{2}$.
  The fluxes at the endpoints ($q_1 = 0$ and $q_1 = N$) can be directly
  computed from the fields there, without any interpolation.
  From the fluxes at the cell ends we construct a numerical flux
  as usual from data in this cell and the neighboring cell in the
  adjacent element.
  From the numerical fluxes at the cell ends we then obtain flux derivatives
  at the cell centers. For all interior cells the centers coincide
  with grid points, so that we immediately obtain flux derivatives at the
  grid points. However, the cell centers near the left and right end
  (marked by the \textcolor{red}{$\times$}) do not coincide with grid points.
  To obtain flux derivatives at grid points $q_1 = 0$ and $q_1 = N$
  we use a weighted average of the flux derivatives at the two closest
  cell centers of this element.
}
\end{figure*}


Figure \ref{FVWENO} illustrates what type of interpolation we use
to obtain values at the midpoints.
Note however, that
$
\left[
\frac{d(J \sqrt{\bar{\gamma}^{\bar{1}\bar{1}}} n^{\bar{1}}_i f^i)}
     {dx^{\bar{1}}}\right]_{\frac{1}{4}}
$
corresponds to a derivative term at a point that is a quarter of a grid
spacing away from the point at $q_1=0$. To obtain a value at
$q_1=0$ we can simply copy this value to the point at $q_1=0$. However, this
very simple procedure results in an error of $O(h)$ and may thus
spoil second order accuracy. We therefore prefer to obtain the
value at $q_1=0$ from
\ba
\label{fv_divf_extrap}
\left[\frac{d(J \sqrt{\bar{\gamma}^{\bar{1}\bar{1}}} n^{\bar{1}}_i f^i)}
           {dx^{\bar{1}}}\right]_{0} 
&:=&
   w  \left[\frac{d(J \sqrt{\bar{\gamma}^{\bar{1}\bar{1}}} n^{\bar{1}}_i f^i)}
                 {dx^{\bar{1}}}\right]_{\frac{1}{4}} \nonumber\\
&+&(1-w)\left[\frac{d(J \sqrt{\bar{\gamma}^{\bar{1}\bar{1}}} n^{\bar{1}}_i f^i)}
                 {dx^{\bar{1}}}\right]_{1} ,\nonumber  \\
                 && 
\ea
i.e., from a weighted average of values at points $\frac{1}{4}$ and $1$.
We usually choose $w=4/3$ to obtain linear extrapolation from points
at $\frac{1}{4}$ and $1$ to $0$.
Note, however, that this kind of extrapolation has two drawbacks. First,
it is not exactly conservative, but violations in mass conservation
still converge to zero with increasing resolution. Second, extrapolation can
become very inaccurate in low-density regions (e.g. in atmosphere
regions around the star).
We therefore have introduced cutoffs below which we simply use $w=1$.
For the cases where we activate these cutoffs, we set $w=1$ if either
$\rho_0 < 10^{-9} \rho_{0,max}(t=0)$,
or
$\epsilon < 10^{-9} \epsilon_{max}(t=0)$,
or
$\epsilon > 10^{3} \epsilon_{max}(t=0)$,
where $\rho_{0,max}(t=0)$ and $ \epsilon_{max}(t=0)$ are the maximum values
of $\rho_0$ and $\epsilon$ at the initial time.

The extrapolation we have just discussed involves extrapolating flux
derivatives. However, as discussed earlier we also need interpolation (e.g.
WENO or linear) to reconstruct the fluxes at the midpoints. For GRHD
we do this by first interpolating the primitive variables to the midpoints.
Then we convert to conservative variables and use them to compute the
fluxes at the midpoints.

Notice that in the scheme described above, which from here on we will dub as
``WENOZ\ts{2}'' in this paper,
we need to exchange only the
values of the fields at the surface of each element with the neighboring
element, just like in the DG method. This means no extra ghost points are
needed, unlike in more traditional finite volume approaches.
We have devised the above scheme for precisely this reason.
Our approach without extra ghost points should in principle scale better,
not just because it needs to communicate less data, but also because each
element needs to communicate only with the six nearest neighbors on its six
faces. As already noted in~\cite{Bugner2018a},
the ghost point approach
in~\cite{Deppe:2021ada,Deppe:2021bhi} needs communication with all 26
neighbors, as soon as patches that are not cuboids are used, e.g.~when cubed
spheres are used.
The price we pay for this is that our scheme is only second order accurate
near the element boundaries.


\subsection{Time integration}
\label{time-int}
Note that Eqs.~(\ref{Integ4}), (\ref{IntegA}), and (\ref{FV-scheme})
still contain time
derivatives, since up to this point we have only discretized spatial
derivatives. This means that these equations represent a set of coupled
ordinary differential equations for the fields $u_{q_1 q_2 q_3}$ at
the grid points. In this paper, we use a standard strongly stability-preserving
third-order Runge-Kutta (RK) time integration method~\cite{Gottlieb2001}
to find the solution of these ordinary differential equations (ODEs).
Such Runge-Kutta methods are only stable when the
Courant-Friedrichs-Lewy (CFL) condition is satisfied, 
i.e., if the time step $\Delta t$ satisfies
\be
\label{cfl}
\Delta t \leq \Delta x_\mathrm{min} / v,
\ee
where $\Delta x_\mathrm{min}$ is the distance between the two closest grid points,
and $v$ is a number that needs to be greater than the largest characteristic
speed. For the pure DG simulations in
Secs.~\ref{stableDG} and \ref{perturb} we simply set an appropriate value
for $\Delta t$ itself.
However, in our hybrid DG+FV/FD scheme, we separately evaluate
\be
\label{cflDG}
\Delta t_\mathrm{DG} = \Delta x_\mathrm{min,DG} / 4
\ee
for the DG elements, and
\be
\label{cflFV}
\Delta t_\mathrm{FV/FD} = \Delta x_\mathrm{min,FV/FD} / 8
\ee
for the FV/FD elements.
Then, the actual time step is set as
$\Delta t = \min\left( \Delta t_\mathrm{DG},  \Delta t_\mathrm{FV/FD}\right)$,
which is used as the time step for all the elements throughout the mesh.


\subsection{Filtering for improved stability}
\label{filters}
Even when the time step satisfies the CFL condition,
aliasing-driven instabilities can still
occur in some cases in the DG elements, e.g., on non-Cartesian patches.
To combat such instabilities, we filter out high-frequency modes in the evolved
fields in the DG elements.
This is achieved by first computing the coefficients $c_{l_1 l_2 l_3}$ in
the expansion
\be
u(\bar{x})J
= \sum_{l_1=0}^N\sum_{l_2=0}^N\sum_{l_3=0}^N c_{l_1 l_2 l_3}
   P_{l_1}(x^{\bar{1}}) P_{l_2}(x^{\bar{2}}) P_{l_3}(x^{\bar{3}})
\ee
of each field $u$, where the $P_{l}(\bar{x})$ are Legendre polynomials,
and
$J = \left|\det\left(\frac{\partial x^i}{\partial x^{\bar{i}}}\right)\right|$
is the Jacobian.
The coefficients are then replaced by
\be
\label{exp_filter}
c_{l_1 l_2 l_3} \to
c_{l_1 l_2 l_3} e^{-\alpha_\mathrm{f} (l_1/N)^s} e^{-\alpha_\mathrm{f} (l_2/N)^s}
                e^{-\alpha_\mathrm{f} (l_3/N)^s} ,
\ee
and $u$ is recomputed using these new coefficients. The values we use
for $\alpha_f$ and $s$ are stated in each of the simulation scenarios in
Sec.~\ref{results}. All these cases result in the coefficients with the
highest $l = N$ to be practically set to zero,
while all others are mostly unchanged.
For the FV/FD elements, we never use filters and instead rely on
Kreiss-Oliger dissipation.


\subsection{Treatment of low-density regions}

In regions of low density, the hydrodynamic fields can develop unphysical
values. This is due to numerical errors during RK substeps, and also due
to the difficulties in recovering the primitive variables from the evolved
conserved variables.
To deal with these problems and to keep the fields in a physically sound state,
we introduce two hydrodynamic numerical mechanisms. We introduce an
artificial atmosphere in the vacuum and low-density regions.
Additionally, we apply some positivity limiters to the fields to further ensure
physicality.


\subsubsection{Atmosphere treatment:}
\label{atmo}

For atmosphere treatment, we consider some density levels as
$\rho_\mathrm{atmo,cut} = f_\mathrm{atmo,cut} \cdot \rho_{0,\mathrm{max}}(t=0)$,
$\rho_\mathrm{atmo,level} = f_\mathrm{atmo,level}
\cdot \rho_{0,\mathrm{max}}(t=0)$, and $\rho_{\mathrm{atmo,max}}
= f_\mathrm{atmo,max} \cdot \rho_\mathrm{atmo,cut}$.
Usually, the
factors $f_\mathrm{atmo,level}$ and $f_\mathrm{atmo,cut}$ are chosen such that the density
levels $\rho_\mathrm{atmo,level}$ and $\rho_\mathrm{atmo,cut}$ are considerably
low compared to the initial maximum density $\rho_{0,\mathrm{max}}(t=0)$ and
$f_\mathrm{atmo,max}$ is set to 1 for most cases.
Having decided the values of both the atmosphere density levels
$\rho_\mathrm{atmo,cut}$ and $\rho_\mathrm{atmo,level}$, we can then determine
the greater of these two, $\rho_{\mathrm{atmo,max}}$, and this becomes our
cutoff or threshold density level for determining whether a point in an
element is in the atmosphere region, or not.

Then, we perform a pointwise
check in each element to see which points, if any, have $\rho_0 <
\rho_{\mathrm{atmo,max}}$. Among the points that have $\rho_0 <
\rho_{\mathrm{atmo,max}}$, we further do a secondary evaluation. The points
that further have $\rho_0 < \rho_\mathrm{atmo,cut}$, i.e., where $\rho_0 <
\rho_\mathrm{atmo,cut} ~,~ \rho_{\mathrm{atmo,max}}$, are essentially treated as
being in the counterpart of the ``vacuum'' region of the mesh.
Hence, at these points, we simply
set a static and cold state to the fields, by setting $\rho_0 =
\rho_\mathrm{atmo,level}$, and setting $\epsilon$ and $Wv^i$ to zero.

However,
we treat the points in the intermediate density range of $\rho_\mathrm{atmo,cut}
< \rho_0 < \rho_{\mathrm{atmo,max}}$ with more caution, as these are the
points in the low-density range transitioning from the atmosphere to the bulk
of the matter in the system. Such points are challenging for all numerical
simulation methods.
For these, simply setting $\rho_0 = \rho_\mathrm{atmo,level}$,
$\epsilon = 0$ and $Wv^i = 0$
will cause us to lose
too much velocity and energy from the system, and deviate from the
correct GRHD solution.
At the same time, we have to prevent the particles in the
transitional density region from gaining
arbitrarily large velocity and energy
due to accumulating numerical errors.
For this purpose, we appropriately choose
a maximum cutoff for the magnitude of
$Wv = \ds \sqrt{Wv_i Wv^i}$ in the atmosphere regions as
$(Wv)_\mathrm{atmo,max} = 0.01$, and another cutoff for the $\epsilon$ magnitude
in the atmosphere as $\epsilon_\mathrm{atmo,max} = 0.1$.
Then, in this transition density range $\rho_\mathrm{atmo,cut}
< \rho_0 < \rho_{\mathrm{atmo,max}}$, if at any point we find
$(Wv)^2 > (Wv)^2_\mathrm{atmo,max}$ we scale the $W v^i$ components at that
point as $Wv^i \rightarrow \ds Wv^i \cdot
\sqrt{\frac{(Wv)^2_\mathrm{atmo,max}}{(Wv)^2}}$. Additionally,
if $\epsilon$ goes above
$\epsilon_\mathrm{atmo,max}$, we simply reset $\epsilon$ to
$\epsilon_\mathrm{atmo,max}$. The density $\rho_0$ is not modified
in the transitional region.

Having set the values for $(\rho_0, \epsilon, W v^i)$ at all the points with
$\rho_0 < \rho_{\mathrm{atmo,max}}$ as stated above, all that
remains is to set $P$ appropriately.
We calculate the pressure from the EoS, which is of the
form given in Eq.~(\ref{eos_general}).
Having set all primitive
variables $(\rho_0, \epsilon, P, W v^i)$ at all
the atmosphere points, we then recompute the
conserved variables using Eqs.~(\ref{D-def})-(\ref{Si-def}).
We state our choices for the parameters
$f_\mathrm{atmo,level}$, $f_\mathrm{atmo,cut}$, $(Wv)_\mathrm{atmo,max}$,
and $\epsilon_\mathrm{atmo,max}$
for each of our simulations in Sec.~\ref{results}.


\subsubsection{Positivity limiters:}
\label{sec:lim}

To prevent our hydrodynamical fields from getting unphysical,
we use a number of positivity limiters.
These mainly target low-density regions, but can activate in
any region where unphysicality occurs.
In this context, we consider ``unphysical'' if (i) mass density $D$ becomes
negative, (ii) energy density $\tau$ becomes negative, or (iii) if
$S > D + \tau$, where $S = \sqrt{S_i S^i}$.

For the actual implementation of the limiters, our treatment differs for
the DG and FV/FD elements.
In a DG element, we first check for conditions (i) through (iii) above
at all points in the element.
However, in practice for $D$, we check if it has fallen below
a lower value of $\rho^\mathrm{lim}_\mathrm{atmo,level} = 10^{-12}
\rho_{0,\mathrm{max}}(t=0)$ instead of demanding simple positivity.
If one of the conditions above is triggered at a point in an element,
we scale the respective variable
$u \in (D, \tau, S_i)$ towards its element average using
\be
\label{pos_limiter}
u \rightarrow \bar{u} + \theta_u \cdot (u - \bar{u}).
\ee
Here $\bar{u}$ is the element average of the variable $u$, with
$0 \leq \theta_u \leq 1$.
For both $D$ and $\tau$, we set the scaling factor as
\be
\label{thDT}
\theta_u = \frac{\bar{u} - u^\mathrm{lim}}{\bar{u} - u_\mathrm{min}}
\ee
where $u_\mathrm{min}$ is the element minimum for the variable $u$,
while $u^\mathrm{lim}$ is the corresponding limiting value,
i.e.~$D^\mathrm{lim} = 10^{-12} \rho_{0,\mathrm{max}}(t=0)$ and
$\tau^\mathrm{lim} = 0$. We then use the $\theta_u$ calculated for 
$D$ and $\tau$ to reset $D$ and $\tau$ in the entire element.
For $S_i$ the scaling factor is
\be
\label{thS}
\theta_{S} = \min\left(
                       \frac{D + \tau - \bar{S}}{S - \bar{S}} - 10^{-14}
                       \right) ,
\ee
where $S = \ds \sqrt{S_i S^i}$ and the $10^{-14}$ subtraction ensures that
$S$ remains below $D + \tau$. The $\min$ in Eq.~(\ref{thS}) refers to the
minimum over all points in the element.
We use $\theta_{S}$ for scaling all the $S_i$ values in the element.
For $D$ and $\tau$ related limiting,
if the denominator becomes $0$, we simply set $\theta_u = 1$, and for $S_i$
we set it to $1 - 10^{-14}$.
Lastly, $\theta_u$ is always constrained within the range $0 \leq \theta_u
\leq 1$.
Note that all these limiting
strategies preserve the element averages of the variables,
which ensure better accuracy for DG elements.
However, if the average itself violates the condition we try to enforce,
the above scaling towards the average is not sufficient to obtain
a physical state. In order to address this eventuality, we also
perform a pointwise limiting for all elements to fix the state of
the fields in any problematic point.
The checked conditions still remain the same as (i) through
(iii) as stated above. In the pointwise limiting, in case at a point $S > D +
\tau$, we simply scale $S_i$ as $S_i \to \ds S_i \cdot
\left| \frac{D + \tau}{S} \right| \cdot f_S^\mathrm{lim}$, where we choose
$f_S^\mathrm{lim}$ appropriately to be $0.99$.
In case we find $D < 0$ or $D + \tau \leq 0$,
we simply set $S_i = 0$. As for $D$, if at any point we find
$D < 10^{-12} \rho_{0,\mathrm{max}}(t=0)$, we simply set
$D = 10^{-12} \rho_{0,\mathrm{max}}(t=0)$.
Similarly, for $\tau$, if at any
point we find $\tau < 0$, we set $\tau = 0$.

For FV/FD elements we only perform the pointwise limiting explained above.

In addition to limiting the variables $(D, \tau, S_i)$ in an attempt to retain
physicality, we also collect information regarding the field conditions in
the nonatmosphere regions during each of our time evolution substeps. This is
because if the unphysicalities arise at any point in an element and limiting
becomes necessary, it is treated as sign of ``trouble'' in an element as the
matter fields are not well behaved in it.
This information is then used in our list of criteria for determining
trouble in an element discussed in Sec.~\ref{fvdg}.

We state our specific parameter choices for
$f_S^\mathrm{lim}$ for our simulations in the next Sec.~\ref{results}.


\subsection{Conversion between conservative and primitive variables}
\label{cons2prim}

We need to convert from the conserved variables $(D, \tau, S_i)$ to the
primitive variables $(\rho, \epsilon, P, Wvi^i)$ at every Runge-Kutta
substep. This needs to happen because we need the primitive variables
for computing flux vectors from Eq.~(\ref{GRHDfluxes}), for setting source
terms from Eq.~(\ref{GRHDsources}), and the atmosphere treatment
from Sec.~\ref{atmo}.
Computing the conserved variables from the primitive ones is straightforward
through Eqs.~(\ref{D-def})-(\ref{Si-def}), but the reverse operation involves
a root finder since the equations cannot be inverted analytically.
The basic principle of conversion
between primitive and conservative variables for the simulations in this paper
remains similar to Sec.~4.3.1 of~\cite{Tichy:2022hpa}, with some
small adjustments. We still use the same procedure described
in~\cite{Tichy:2022hpa} to solve for the root value of
\be
Wv := \sqrt{Wv^i Wv_i}
\ee
with a root finder, as in Appendix C of~\cite{Galeazzi:2013mia}, from the
function
\be
f(Wv) = Wv - \frac{\sqrt{S_i S^i}}{D h(Wv)} .
\ee
However, unlike in~\cite{Tichy:2022hpa}, first, our current procedure takes
into account whether the point we are solving for is in the atmosphere region
or not.
If at any point we find $D < \rho_\mathrm{atmo,cut}$,
we simply set $\rho_0 = D$ and
$\epsilon = P = Wv^i = 0$, without going through the actual root solving
procedure. This is different from the procedure in~\cite{Tichy:2022hpa} as
that method would perform the conversion everywhere through the root solver.
By avoiding root solving in these low-density regions, which is mostly the
regions that result in unphysical states as mentioned before, we avoid the
associated numerical complications that might cause the root solver to fail.
Additionally, even when we do perform the actual root solving, we take certain
precautions against possible complications. First, as root finding can
fail due to round off error near the upper and lower limits of the interval
in which the root must lie according to~\cite{Galeazzi:2013mia},
we extend the maximum of the range to $1.1 (Wv)_\mathrm{max}$,
with $(Wv)_\mathrm{max}$ being the upper limit. Secondly, we set a floor
value for $\epsilon_\mathrm{floor}$ such that if $\epsilon$ falls below this
through the conversion process, we simply set $\epsilon =
\epsilon_\mathrm{floor}$. The particular value for $\epsilon_\mathrm{floor}$
used is stated for the simulations in Sec.~\ref{results}.
The last point where we deviate from the prescription
in~\cite{Tichy:2022hpa} is that we do not perform any kind of rescaling if
$Z > Wv$, where
$Z = \ds \sqrt{S^i S_i}/ (W h \rho_0)$.
Just like for the limiters, we also collect information
about any trouble when converting from conserved to primitive variables
at each point in an element during an evolution substep.
The reasoning behind this is that in
an element that has smooth and well-behaved matter fields, the conversion
procedure should be carried out seamlessly. However, any kind of problem or
difficulty during the conversion process will indicate that the matter fields
are not well behaved, and hence the element is marked as troubled.
This information is then used in our trouble  criteria list as mentioned
in Sec.~\ref{fvdg}.


\subsection{Hybrid DG+FV/FD evolution scheme}
\label{fvdg}

We want to use DG in the elements that are not troubled.
In the troubled elements, we want to use FV to
evolve the matter fields and FD to evolve the gravity
fields.
The concept of ``trouble'' is defined mainly based on the smoothness
of the matter fields in an element. If the matter fields in an element are not
smooth, the element is deemed ``troubled''.
Additionally, an element is also considered
``troubled'' if unphysicalities occur in the matter fields in the element
either in low-density regions or during conversion from conserved
to primitive hydrodynamic variables.
For determining
trouble in an element, we use the following prescription:

i.
We perform a full-time evolution step from $t$ to $t + \Delta t$ through the
RK time evolution scheme described
in Sec.~\ref{time-int} and at each RK substep,
we collect information about troubles during the substeps.
These include nonphysical behavior that triggers one of the
positivity limiters discussed in Sec.~\ref{sec:lim},
and problems when converting between conserved and primitive variables
as described in Sec.~\ref{cons2prim}.

ii.
At the end of each full RK step, we check each element for
any indication of trouble
using the following list of criteria exactly in this order,
which then constitutes our troubled-element indicator (TEI):

	\begin{enumerate}
	\item
	Look at the conserved density variable $D$ using two
	cutoff densities:
	\begin{align}
    D &\leq& \rho_\mathrm{cut,low} & & & & & &
		                            & \rightarrow & \mathrm{all~ good},\nonumber \\
		  &   &  \rho_\mathrm{cut,low}  
              &\leq& D
                  	&\leq& \rho_\mathrm{cut,hi} & & 
                              & \rightarrow & \mathrm{troubled},
													 \nonumber\\
    & & & & & &              \rho_\mathrm{cut,hi}  
                                  &\leq& D 
                                      &\rightarrow & \mathrm{neutral} 
                                      \label{TroubleD}
  \end{align}

  The values of $\rho_\mathrm{cut,low}$ and
  $\rho_\mathrm{cut,hi}$ are set as
  $\rho_\mathrm{cut,low} = f_\mathrm{cut,low} \cdot \rho_{0,max}(t=0)$ and
  $\rho_\mathrm{cut,hi} = f_\mathrm{cut,hi} \cdot  \rho_\mathrm{cut,low} $.
  If the result is neutral, criteria below are checked.
  
  \item
	Look at data about any trouble collected during the evolution step.
	\item
	Look at the relaxed discrete maximum principle
	(RDMP)~\cite{Dumbser2014a,Deppe:2021ada},
	for the $(n+1)$th time step solution $ u^\star (t^{n+1}) $ and demand that
  \be
    \label{rdmp}
		\ds \min_{E_n}[u(t^n)] - \delta  < u^\star (t^{n+1})
                                                  < \max_{E_n}[u(t^n)] + \delta
	\ee
	with $  \delta = \max \left( 10^{-7} , 10^{-3}
	       \left\{ \ds \max_{E_n}[u(t^n)] - \min_{E_n}[u(t^n)]  \right\} \right)$,
	where $E_n$ denotes nearest neighbors (i.e.~von Neumann neighbors)
	of the element plus the element itself. If this condition does not
	hold in an element, then the element is considered troubled.

  \item
  Attempt a conversion from the conserved to primitive variables
  for the upcoming time step in the element.
  If the attempt fails, consider the element as troubled.

	\item
  Look at the expansion coefficients of the fields
  in the basis polynomials and decide
	based on Persson method~\cite{Persson2006a} in an element by demanding
	\be
    \label{persson}
		\ln\left( \frac{c^2_\mathrm{hi}}{c^2_\mathrm{all}} \right)
			< - \alpha_N^{\mathrm{evo}}
      \ln\left( \max(N_{1,\mathrm{uf}}, N_{2,\mathrm{uf}},
                            N_{3,\mathrm{uf}}) \right) ,
	\ee
	where
	\ba
		 c^2_\mathrm{hi} &=&
		    \ds \sum_{l_1 = 0}^{N_{1,\mathrm{uf}}}
        \ds \sum_{l_2 = 0}^{N_{2,\mathrm{uf}}} 
    c_{l_1 l_2 N_{3,\mathrm{uf}}}^{2}
		  +	\ds \sum_{l_1 = 0}^{N_{1,\mathrm{uf}}}
        \ds \sum_{l_3 = 0}^{N_{3,\mathrm{uf}}}
    c_{l_1 N_{2,\mathrm{uf}} l_3}^{ 2}  \nonumber \\
    & & + \ds \sum_{l_2 = 0}^{N_{2,\mathrm{uf}}}
         \ds \sum_{l_3 = 0}^{N_{3,\mathrm{uf}}}
    c_{N_{1,\mathrm{uf}} l_2 l_3}^{ 2},
                                                              \label{ncsurf}
	\ea
	and
	\be
		c^2_\mathrm{all} =
        \ds \sum_{l_1 = 0}^{N_{1,\mathrm{uf}}}
        \ds \sum_{l_2 = 0}^{N_{2,\mathrm{uf}}}
        \ds \sum_{l_3 = 0}^{N_{3,\mathrm{uf}}}c_{l_1 l_2 l_3}^2 .
																															\label{ncvol}
	\ee
  Violation of condition (\ref{persson}) is considered indication of trouble.
  When considering this criterion, we have to account for the fact that
  we apply exponential filters to our matter field coefficients in the DG
  elements, as per the filtering description in Sec.~\ref{filters}.

  Hence,
  we should consider only the $N_{l,\mathrm{uf}}$ number of
  coefficients that remain unfiltered for each direction, as these are
  the ones that truly represent the smoothness of the fields in an element.
  In order to determine which coefficients have not been filtered, we only count
  those where the filter factor is above some factor $f_\mathrm{unfiltered}$,
  i.e., we only include $c_{l_1 l_2 l_3}$ if
  $c_{l_1 l_2 l_3} \geq
   f_\mathrm{unfiltered} c_{l_1 l_2 l_3}^\mathrm{unfiltered}$.
  For the choice of filtering in the
  simulations presented in Sec.~\ref{results}, we find that the number of
  unfiltered coefficients $N_{l,\mathrm{uf}}$
  is related to the number of total
  coefficients $N_l$ in each direction as $N_{l,\mathrm{uf}} = N_l - 1$.
  Note that the coefficients are always computed using the DG grid points.
  This is straightforward in an element using the DG scheme, as we have
  the values of the variable we are checking already at the DG grid points.
  However, if we want to check the criterion in a FV/FD element, we first
  interpolate onto the DG grid points using a twelfth-order Lagrange
  interpolation scheme, and then compute the expansion coefficients
  on this reconstructed DG grid.
  Lastly, $\alpha_N^{\mathrm{evo}}$ is a relaxation parameter.
  For our simulations, we set $\alpha_N^{\mathrm{evo}} = 8$ in the DG elements
  and $\alpha_N^{\mathrm{evo}} = 9$ in the FV/FD elements, except for
  the blast wave propagation tests in Sec.~\ref{srhdtestid}. We
  relax our parameters in  Sec.~\ref{srhdtestid}, by using $4$ for DG
  elements and $5$ for FV/FD elements.
  The higher values in FV/FD elements is because we want
  to be more cautious when switching from FV/FD to DG as DG will not work
  well in a troubled element.
  From our experience, this Persson trouble indicator
  plays an especially crucial role in the accuracy of the simulation.
	\end{enumerate}

These  criteria are checked in order,
and if any one of them indicates trouble at any point
in an element, the element is marked
as troubled by the TEI and no further check is performed.

iii.
If an element is marked troubled by the TEI,
then if we are using DG, we switch to FV/FD and redo the full RK
step (not just the most recent substep).
Conversely, if we are using FV/FD then we keep doing FV/FD for the next RK step.
On the other hand, if the element is not marked troubled,
then if we are using DG, we try doing DG for the next step.
However, if we are using FV/FD, then we switch to DG only after
11 consecutive time steps with no trouble to be safe. In the DG elements,
we use half the number of points as in the FV/FD elements, as DG can
generate higher accuracy results at low computational cost.


\section{Numerical results}
\label{results}

Given all three of our evolution schemes, namely the DG
with the hydrodynamic fields coupled to a dynamic metric,
the modified FV/FD scheme and the hybrid DG+FV/FD scheme, are new
implementations in the \nmesh~program, it is necessary
to ensure the proper functioning of these methods. Although the
DG scheme had been tested somewhat in our previous paper~\cite{Tichy:2022hpa},
the tests were performed separately for the gravity evolution system
(through single black hole evolutions)
and the matter evolution system (through neutron star evolutions
under the Cowling approximation of a static metric~\cite{Cowling1941}).

\begin{table*}
	\caption{\label{NSID} Specifics of the different
	single neutron stars we simulate. The first column states
	the name and the abbreviation we assign.
	The second column states the initial central density.
	The third and fourth column state the rest mass and the initial star
        radius (in Schwarzschild coordinates).
	The fifth column states the strength of the initial
	perturbation (either in pressure $P$ or density $\rho_0$)
	and the sixth column states the speed imparted to
	the CoM. The last column states the choices
	of evolution scheme we use to simulate the case. In each case,
	we use the ideal EoS of Eq.~(\ref{eos}) with $\kappa = 100$ and $n = 1$.
	}
	\centering
	\begin{tabular}{c c c c c c c}
	\hline
	\hline
	Name & $\rho_{0,\mathrm{C}} \cdot 10^3 M_\odot^2$ & $m_0/M_\odot$ & $R_\mathrm{Sch}
	 / M_\odot$
	          					& $\lambda$ & $v_\mathrm{CoM}$ & Scheme(s) \\
	\hline
	\hline
	Stable (S) & 1.28 & 1.5061762 & 9.5856240 & 0 & 0 & DG, DG+FV/FD\\
	Perturbed (P) & 1.28 & 1.4911824 & 9.5856240
                      & 0.05 [Eq.~(\ref{pertP})]   & 0 &    DG \\ 
	Unstable (U) & 7.99 & 1.5353840 & 5.8380336 
                      & 0,-0.01 [Eq.~(\ref{pertrho0})] & 0 & FV/FD, \\
	Boosted (B) & 1.28 & 1.5061762 & 9.5856240 & 0 & 0.1, 0.5 & DG+FV/FD \\
	\hline
	\hline
	\end{tabular}
\end{table*}
In this paper we present simulations that we have
performed to ensure the robustness and stability of our new methods.
The results of these simulations are presented here through a number of plots;
the data related to them can be found in~\cite{Adhikari2025b}.
We start with a scalar wave test, as this helps us test our
new DG+FV/FD scheme purely by itself, without the numerical complexities
arising from matter fields in neutron star systems.
Having checked proper convergence for this simple case,
we move on to 
evolving initial data with
special relativistic Riemann discontinuities in the initial data
to test hydrodynamic shock propagation.
Next, we simulate single neutron stars.
We simulate four different cases of single neutron star
initial data that are outlined in Table~\ref{NSID}, while the resolutions
used for each of the initial setup are described in Table~\ref{NSres}.
\begin{table*}
	\caption{\label{NSres} Initial resolution, number 
	and type of elements across the star for different
	single neutron star cases we simulate. The first column is the name of the case
	as per Table \ref{NSID} and the second column is the corresponding
	section where the case is discussed. $L$ in the third column states
	the dimension of the central cube. The fourth column states the evolution
	scheme or schemes used. The fifth and sixth columns state the number
	of points in each direction in DG and FV/FD elements respectively,
	as applicable. The seventh column states center velocity $v_\mathrm{CoM}$.
	The eighth column specifies the highest level of h-refinement
	in the refinement region centered on the star. The last column states the
	initial total number of elements across the star, while specifying
	how many are of each type.
	When counting these elements, we also
	count those which partially contain the star at its surface.
	In the DG+FV/FD cases, the FV/FD elements are on the star surface, while
	DG elements are in the center. The elements for the $v_\mathrm{CoM} = 0.5$
	case are counted parallel ($ \parallel$) to $\vec{v}_\mathrm{CoM}$ direction
	and perpendicular ($\perp$) to $\vec{v}_\mathrm{CoM}$ direction.
	}
	\centering
	\begin{tabular}{c c c c c c c c c}
	\hline
	\hline
	Name & Sec.  & $L$ & Scheme & $n$(DG) & $n$(FV/FD) & $v_\mathrm{CoM}$
																	  & $l$(max) &  Star elements \\
	\hline
	\hline
	(S)  & \ref{stableDG} & $32$ & DG & $5$ & - & 0 & $4$ & $10$ DG\\ 
			&			 &  &    &  &  &  & $5$ & $18$ DG \\ 
	    &   	 &  &    &  &  &  & $6$ & $34$ DG \\ \cline{2-9}
			& \ref{stablehy} & $52$ & DG+FV/FD  & $8$ & $16$
																	& 0 & $5$ & $8$ DG $+$ $4$ FV/FD
	\\ \hline
	(P) &\ref{perturb} & $32$ & DG & $5$ & - & 0 & $4$ & $10$ DG\\
			&			 & 		&    &  &  &  & $5$ & $18$ DG \\
			&			&	&    &  &  &  & $6$ & $34$  DG
	\\ \hline
	(U) & \ref{fvmigpref} & $16$ & FV/FD & - & $12$,$18$,$28$,$42$
																			& 0 & $3$ & $5$ FV/FD \\  \cline{2-9}
	    & \ref{mighref} & $16$ & FV/FD & - & $16$ & 0 & $3$ & $5$ FV/FD \\
	    & &       &       &  &  &  & $4$ & $9$ FV/FD \\
			&	&			 &       &  &  &  & $5$ & $17$ FV/FD \\  \cline{2-9}
			& \ref{mighy} & $16$ & DG+ FV/FD & $8$ & $16$ & 0 & $3$ & $5$ FV/FD \\
			&				 &       &  &  &  & & $4$ & $9$ FV/FD 
	\\ \hline
	(B) & \ref{boost} & $52$ & DG+FV/FD & $8$ & $16$
																& $0.5$ & $5$ & $5$ DG $+$ $2$ FV/FD
																		$ \parallel \vec{v}_\mathrm{CoM}$, \\
			&			 &  &   &  &         	&  &  & $5$ DG $+$ $4$ FV/FD
																				$ \perp \vec{v}_\mathrm{CoM}$
			\\ \cline{7-9}
			&	     &       &	         &   &  
                                  & $0.1$ & $4$ & $1$ DG $+$ $4$ FV/FD \\
			&	     &       &	         &   &    &  
                                  & $5$ & $5$ DG $+$ $4$ FV/FD
	\\
	\hline
	\hline
	\end{tabular}
\end{table*}
Each of these cases
are discussed in detail in the sections below. Out of these, (S) tests
for stability and robustness, while the rest test the ability
of the program to handle various types of dynamic scenarios.
In order to study the behavior of the numerical results with
increasing resolution, we use both p-refinement and h-refinement.
P-refining an element simply involves increasing the number of
points in the element in each direction as necessary. In case of h-refinement
of an element, we take the element and split it in two in each direction.
This results in eight new elements replacing the old element.
Each of these eight new elements
contains the same arrangement of points as the parent element. For DG elements,
p-refinement increases the basis polynomial order, which would cause
an exponential dropoff in the error in the results in the case of smooth
fields, but can also prove problematic when nonsmooth fields are present
in an element,
as these can give rise to sharp noisy features in the higher basis modes.
In this case, for DG elements, it is
more convenient to increase resolution using h-refinemet instead.
For the case of pure FV/FD
evolutions on the other hand, we can employ both
p-refinement and h-refinement as necessary.

To the best of our knowledge, this is the first time DG methods have been
used for boosted neutron stars (B) and for unstable neutron stars (U) (that
migrate to a stable state) in fully 3D simulations.
While~\cite{Deppe:2021ada,Deppe:2021bhi, Deppe:2024ckt} evolve single TOV
stars with a hybrid DG-FD scheme, they study the initially static
configuration and a rotating configuration. Note, however,
that~\cite{Deppe:2024ckt} also has successfully
evolved a binary inspiral and merger in full 3D without any symmetry
assumptions, which is a case we have not studied yet with \nmesh.
Although simulations of the
initially unstable configuration are performed with DG methods
in~\cite{Radice:2011qr}, a 1D spherically symmetric implementation is used.
This case setup is also studied in~\cite{Bugner2018a}, where simulations are
performed using spherically symmetric and also axisymmetric cartoon
methods~\cite{Alcubierre99a,Pretorius:2004jg}. As such, none of the prior
simulations are in full 3D. Additionally, none of them evolve boosted
neutron stars.

\subsection{Tests with a smooth scalar field}
\label{scalar}

In order to keep message passing interface (MPI) communication at a minimum, our particular FV method
does not use any extra ghost points (unlike more traditional finite volume
codes). Communication has to happen only for points on the element surfaces
as in the DG method. This is crucial for good strong scaling. On the other
hand, our FV method becomes less accurate near element boundaries, because
we need to drop to a lower order FV scheme there, due to the lack of ghost
points. We have tested the method by evolving a smooth scalar field,
using the same conservative evolution equations as described in Sec.~4.1
of~\cite{Tichy:2022hpa}.
We choose the mesh to be a 2-dimensional square that is equally divided
into nine smaller square patches.
In the middle patch we use FV, while the surrounding eight patches use 
DG.
We then uniformly h-refine each patch, so that each of the nine patches is
covered by an equal number of elements. In each element we use only four points
(in each of the two directions), which is a challenging test for our
FV method that works best if the number of points in the interior is larger
than the number of element boundary points, because our method falls back to
lower accuracy at element boundaries as explained in Sec.~\ref{modFV}.
\begin{figure}
\centering           
\includegraphics[width=\linewidth]{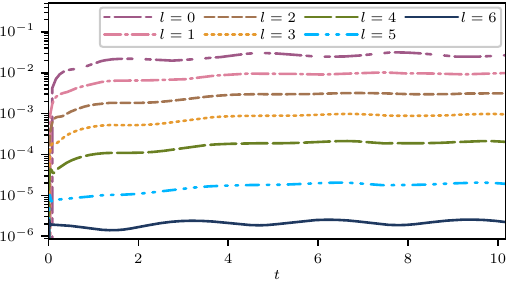}
\caption{            
\label{scalarDGFVconv} 
\small{This figure shows the L2 norm of the error in the scalar field for
a mesh that contains both DG and FV elements. We show the results for
seven resolutions corresponding to h-refinement levels of
$l = 0, 1, 2, 3, 4, 5, 6$ and $7$. The error is consistent with 2nd 
order convergence.}
}
\end{figure}
In Fig.~\ref{scalarDGFVconv} we show
how the numerical error converges to zero with resolution. Our
method is consistent with second order convergence.
While second-order convergence may not sound too impressive, we note that in
shocked regions we cannot expect more than first-order convergence in any
case. Thus, our method is adequate when used in elements with
shocks. Our plan is then to use this FV method only in the elements where
the matter is not smooth, while we plan to use the DG method everywhere
else. In this way we should get the benefits of both FV and DG as suggested
in~\cite{Deppe:2021ada,Deppe:2021bhi}.

\subsection{Special relativistic blast wave tests}
\label{srhdtestid}

As a next step, we propagated a so-called special relativistic blast
wave in one dimension. The initial setup
has discontinuities in the matter fields of $\rho_0$ and
$P$, as in a Riemann problem. We study the case from~\cite{Marti99}
with $(\rho_0, P, v_x)$ given by
\be
\label{srhdID}
(\rho_0, P, v_x) = \left\{
\begin{array}{ll}
	(10.00, 13.33, 0.00)  & x\leq 0.5,  \\
	(1.00, 0.00, 0.00)    & x > 0.5.
\end{array}
\right.
\ee
We also use the analytic
solution from the same article to check the accuracy of our simulation
result. We then evolve the initial discontinuity from Eq.~(\ref{srhdID})
using the full GRHD equations
on a fixed Minkowski metric.
For these tests, the time step was set according to the prescription in
Sec.~\ref{time-int}. 
\begin{figure}
\centering           
\includegraphics[width=\linewidth]{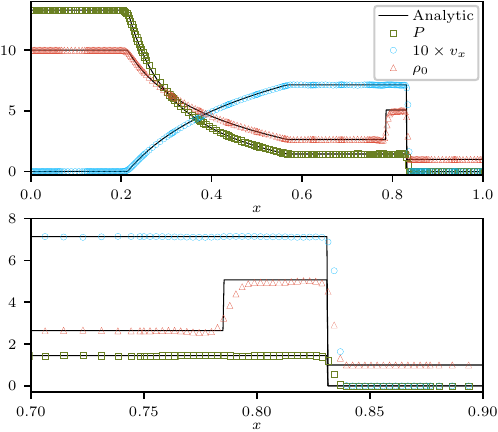}
\caption{            
\label{srhdprof} 
The plot shows blast wave profiles for pressure $P$,
rest-mass density $\rho_0$ and speed $v_x$ (times 10)
at $t = 0.4$ after evolving the initial shocks in $P$ and $\rho_0$.
}
\end{figure}
We set up the mesh with $16$ adjacent Cartesian elements
along the $x$-axis. Initially, we set the element at the left
boundary and the middle element containing the initial discontinuity
to FV/FD containing $24$ equally spaced grid points,
and the rest of the elements to DG, with $12$ grid points.
During the evolution, we use our TEI from
Sec.~\ref{fvdg} to decide if an element should use the DG or FV/FD method.
For this case, we switch off the criterion
from Eq.~(\ref{TroubleD}),
as the purpose of it is to handle
low-density regions present in neutron star simulations, which is not
relevant in this simulation.
We also switch off the RDMP criterion,
as it activates in regions where the analytic solution has zero slope, due
to small amplitude variations in the numerical solution.
These small amplitude variations can also
trigger the Persson indicator if not relaxed enough.
So, for the condition from Eq.~(\ref{persson}),
we choose $\alpha_N^\mathrm{evo} = 4$ for the DG elements and
$\alpha_N^\mathrm{evo} = 5$ for the FV/FD elements. These values are
lower than what we use for our star simulations in
Secs.~\ref{stablehy}, \ref{mighy} and \ref{boost}, where we choose
more conservative ones that result in more elements using FV/FD.

In Fig.~(\ref{srhdprof}) we plot the primitive variables. We show the
entire mesh (top plot) as well as the contact discontinuity and
the shock front (bottom plot), after evolving until time $t = 0.4$.
We see that the elements near the nonsmooth
regions evolve with FV/FD, while most other
elements evolve with DG, as can be seen from the top plot.
In both plots, we see that the fields in our simulation agree very closely
with the analytic solutions that are plotted as solid black lines.
This demonstrates that our hybrid DG+FV/FD method can successfully handle
evolutions of shock fronts in matter fields, which is a crucial initial
test, as such hydrodynamic shocks are known to develop during
late inspiral and merger phases of a binary neutron star merger.

\subsection{Stable configuration (S)}

This first case (S) starts with initial data for a static
TOV configuration
for testing stability and robustness of our program.
Although we start with a static star, it gets perturbed through numerical
errors throughout the duration of the simulations.
Case (S) was simulated successfully with pure DG as well as DG+FV/FD.
In (S), for setting up the initial data, we use a polytropic EoS which
has the form in Eq.~(\ref{inidataeos})
while we evolve with the ``gamma-law'' EoS, which is of the form in
Eq.~(\ref{eos}) that allows for heating.
We use this EoS for all the neutron star simulations of all the setups in
Table~\ref{NSID}. Similarly, for all the
neutron star simulations in this paper, we use the Rusanov
or local Lax-Friedrichs (LLF) flux for the evolution of the 
hydrodynamic (hydro) system, while we use the so-called ``upwind'' flux
for the metric evolution with the GH system. Both of these numerical flux
implementations are the same as described in Sec.~2.3
of~\cite{Tichy:2022hpa}.

\subsubsection{Pure DG (S) simulations}
\label{stableDG}

\begin{figure}
	\includegraphics[width=\linewidth]{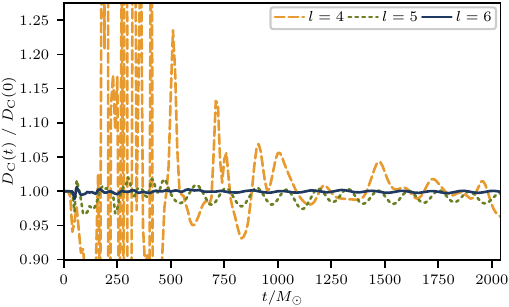}
	\caption{\label{dgStableDpt0} This figure shows central density
	$D_\mathrm{C}$ for (S) with pure DG simulations corresponding to
	h-refinement levels of $l = 4, 5$ and $6$, normalized to the initial
	value. Large amplitude Gibbs oscillations originating from the star
	surface spread across it generating large amounts of
	noise. This is especially true near the beginning, but continues to plague
	the lowest resolutions throughout. As resolution increases, the star settles
	down into more systematic oscillations consistent with the star's
	fundamental oscillation frequency.}
\end{figure}

For the DG case, we set up the mesh using a cubical patch at the center,
surrounded by six cubed sphere patches on all sides, which in turn have
another layer of six cubed sphere patches
forming a spherical shell covering them, resulting in total 13 patches.
We then h-refined the central cubical patch successively
four, five and six times, while refining all the
other patches one time less in all the cases,
resulting in $10240$, $81920$, and $655360$ total elements after h-refining.
In each of these elements, we use $5 \times 5 \times 5$ grid points.
The star is contained in the central cube with side length $32$,
which is centered on the star, while the outermost boundary of the
entire mesh is at 320. The resultant initial number of elements across
the star for the $l=4, 5$, and $6$ are stated in Table \ref{NSres}.
We set the time steps for the different
resolutions to be $0.1$, $0.05$ and $0.025$ respectively.
For the artificial atmosphere, we set  $f_\mathrm{atmo,level} = 10^{-12}$,
$f_\mathrm{atmo,cut} = 1.01 \times 10^{-12}$, $f_\mathrm{atmo,max} = 10$,
$\epsilon_\mathrm{atmo,max} = 0.1$, and $(Wv)_\mathrm{atmo,max} = 0.01$
as per the description in Sec.~\ref{atmo}. We also apply our positivity
limiters with elementwise limiting of $D$, $\tau$, and $S_i$,
but not the pointwise fixes described in Sec.~\ref{sec:lim}.
\begin{figure}
	\includegraphics[width=\linewidth]{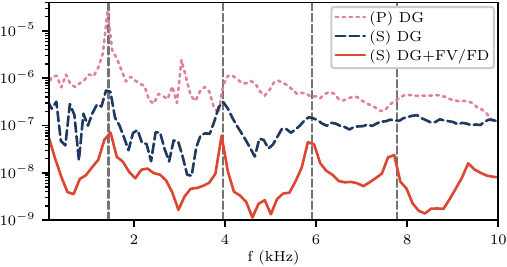}
	\caption{\label{dgDfft}
  This figure shows the Fourier spectrum of the
	central point density $\rho_{0,\mathrm{C}}$ for (S) with both
  pure DG and DG+FV/FD and (P) with pure DG.
	The pure DG (S) and (P) data are taken from the
	simulations with $l = 6$ levels of maximum h-refinement,
  while (S) DG+FV/FD has a maximum of $l = 5$.
  The corresponding dimension of the finest elements resolving the star
  for the pure DG (S) and (P) cases
  is $738.313$ m, while that in the DG+FV/FD (S) case is $2399.516$ m.
  The average grid spacing in this finest element for the pure DG (S) and (P)
  simulations is $\Delta x_\mathrm{avg,DG} = 184.578$ m.
  The DG elements in the interior of the star
  in the DG+FV/FD (S) case have $\Delta x_\mathrm{avg,DG} = 342.788$ m,
  while FV/FD elements on the star surface
  have $\Delta x_\mathrm{FV/FD} = 171.394$ m.
  The dashed
	vertical lines correspond to the expected frequencies for the various
	oscillations modes of the star, as stated in~\cite{Font:2001ew}.
	We see that the frequency for the principal and secondary modes
  from the oscillations matches perfectly with
  the theoretical value~\cite{Font:2001ew} for (S) with both
  pure DG and DG+FV/FD, while even
  the third peak matches for (S) with DG+FV/FD.
	The principal mode is also matching for the pure DG (P) case.}
\end{figure}
This limiter treatment
is essentially the same as that described
in~\cite{Tichy:2022hpa}. We restrict
the lower limit of $D$ to $10^{-12} \rho_{0,\mathrm{max}}(t=0)$ and $\tau$
to $0$, while enforcing $S < D + \tau$, as described when limiting.
For the matter fields, as per Eq.~(\ref{exp_filter}),
the filter uses $s = 32$, while we 
use $s = 40$ for the GH system. For both cases we have $\alpha_f = 36$.
For these simulations, we do not use the conserved to primitive variable
conversion prescription mentioned in Sec.~\ref{cons2prim}. We retain the
method for conversion described in
Sec.~4.3.1 of~\cite{Tichy:2022hpa}. This is
because we performed simulation for the
same physical system successfully under
Cowling approximation
with this method that has been presented
in Sec.~4.3.4 in~\cite{Tichy:2022hpa}.
For the GH system
we set the constraint damping
parameters to be $\gamma_0 = 0.1$,
$\gamma_1 = -1$ and $\gamma_2 = 1$.
The gauge is set such that the gauge
source function is initialized according
to the initial metric and then is kept
constant throughout the simulation, i.e,
we set 
\be
\label{freeze}
H_\alpha(t=0) = -\Gamma_\alpha(t=0), \quad \partial_tH_\alpha=0.
\ee

\begin{figure*}
	\includegraphics[width=\linewidth]{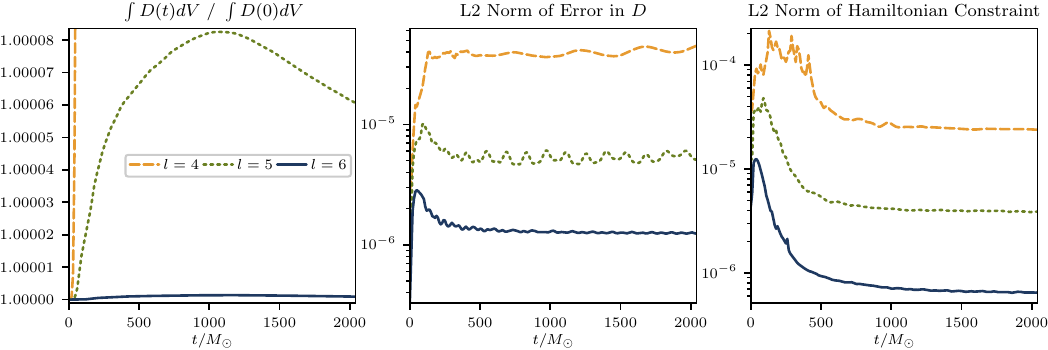}
	\caption{\label{dgStabletData} This plot shows three quantities
  for (S) with pure DG for resolutions $l = 4, 5,$ and $6$:
  total rest mass $\int D dV$ in the entire mesh normalized to the
  initial value (left), L2 norm of error in $D$ (center),
  and L2 norm of the Hamiltonian constraint (right).
  The error quantities (center and right) converge with order $2.6$.
  Mass conservation improves even faster with resolution.
  Here $dV = \sqrt{\gamma} d^3 x$.
  The legend applies to all plots.}
\end{figure*}

The run with
the five times h-refined central cube was evolved stably until
a time of $t = 5000 M_\odot$, even though the plots for this case
only show data until $t = 2000 M_\odot$ time due to high computational costs.
As can be seen from Fig.~\ref{dgStableDpt0},
the central density oscillates
with time, with higher erratic oscillations towards the beginning.
These errors are larger in
the beginning in this pure DG simulation
due to strong Gibbs oscillations arising
from the surface of the star. This is because in the initial setup, the
nonsmooth feature on the surface of the star is at its sharpest,
which produces large Gibbs oscillation in the DG scheme,
as the surface lies in the middle of elements.
These oscillations spread throughout the star and produce strong
numerical disturbances. However, as the simulation progresses, our
stabilizing mechanisms such as the positivity limiters, the filters
and the atmosphere damp the spurious Gibbs oscillations,
which leads to smoother star oscillations.
Hence, at later times, we see a systematic oscillation for the two higher
resolutions, which is in agreement with the spectrum expected from
such a setup, as can be seen from Fig.~\ref{dgDfft},
that shows the Fourier spectrum of the central density $\rho_{0,\mathrm{C}}(t)$
for the case with six central refinement levels for (S) with pure DG.
For this simulation, the dimension of the finest elements covering the
star is $0.5~(738.3125$ m), and the spacings between
the Gau\ss-Lobatto grid points
in these elements are: $\Delta x_\mathrm{min,DG} = 0.0732~(108.123$ m),
$\Delta x_\mathrm{max,DG} = 0.177~(261.033$ m), and
$\Delta x_\mathrm{avg,DG} = 0.125~(184.578$ m),
with the average being the grid spacing
in an element of the same size with same number of equidistant grid points.
In this figure we see that the principal frequency,
as well as the second peak in the spectrum match
the predicted value from literature~\cite{Font:2001ew}.
The higher modes are buried
in numerical noise due to the lack of resolution,
though the third peak can be discerned to be
approximately aligning with the correct value.
Note that the numerical noise in the oscillations decreases with
increasing resolution.
We see the highest resolution is demonstrating consistent and persistent
oscillations at the proper frequency, while for the middle resolution,
they get damped slowly by the filters we apply.
The lowest resolution of four central refinement levels is swamped by
Gibbs oscillations, and is depicted only to show
improvement with resolution.
We next consider mass conservation
by looking at the volume integral of the rest mass density $D$ shown
in the left plot of Fig.~\ref{dgStabletData}.
It is very clear that mass
conservation improves drastically with increasing resolution. The small
amount of discrepancy in the highest resolution is attributed to the fact that
we use positivity limiters that conserve the mass only when the element average
is above the limits imposed. Furthermore, pointwise corrections to $D$ arise
due to the atmosphere treatment. In this case, the $D$ value is pushed up
whenever it wants to fall below the floor value set by us. Hence,
in these situations the integral can increase.

We also study error quantities such as the L2 norm of the error in 
the conserved density $D_\mathrm{C}$ and the Hamiltonian constraint. We look
at these quantities in the central box to get a true representation as
that is where the star is located, while the rest is atmosphere.
These are shown in the center and right plot of Fig.~\ref{dgStabletData}.
As can be seen, these demonstrate convergent behavior with increasing
resolution, with a measured convergence order of $2.6$ for both
$D_\mathrm{C}$ and the Hamiltonian constraint.

\begin{figure}
	\includegraphics[width=\linewidth]{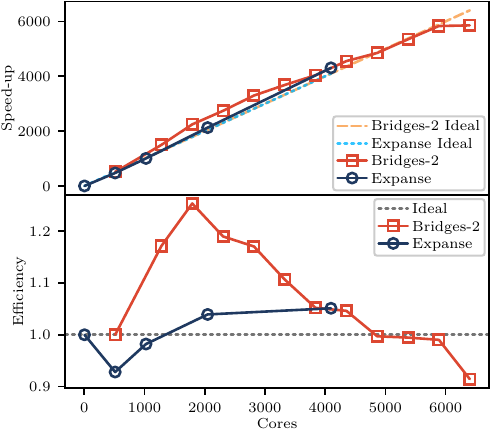}
	\caption{\label{scaling}
  This plot shows speedup and parallel efficiency data for strong scaling tests
	on Bridges-2 and Expanse for the pure DG (S) simulation with $262144$
	elements. The dashed lines in the top plot
  represent ideal speedup curves for the
	range of cores used on each supercomputer, while the solid lines
	represent the actual measured data for \nmesh. As we can see, the ideal
	and real data are in good agreement for speedup, demonstrating strong scaling.
	}
\end{figure}

We have also performed strong scaling tests with this (S) setup with pure DG,
to demonstrate \nmesh's capability to scale efficiently on a large number of
computational cores or processors. This is shown in Fig.~\ref{scaling}.
The top plot shows speedup vs. number of cores used for the run,
while the lower plot shows the parallel efficiency.
Speedup when
using $N$ number of compute cores is defined as $S(N) = t_r(1)/t_r(N)$, where
$t_r(N)$ is the run-time measured for $N$ cores, with each core running
one MPI process.
The parallel efficiency is then defined as $E(N) = S(N)/N$,
which is ideally $1$.
Since a single MPI process cannot obtain enough memory
for the simulations, $t_r(1)$ is estimated from
$t_r(1) = N_{\mathrm{min}} t_r(N_{\mathrm{min}})$, where $N_{\mathrm{min}}$
is for the run with the lowest number of MPI processes performed in each case.

We performed these runs on both the Bridges-2 and Expanse supercomputers.
In our case, we have $N_{\mathrm{min}} = 3$ for the Expanse
supercomputer and $512$ for the Bridges-2 supercomputer,
which is used for the data in Fig.~\ref{scaling}. We go up to $4096$
cores on Expanse and $6400$ cores on Bridges-2, which were the
maximum core count available to us in the respective computers.
The simulation we tested this
with had total $262144$ elements in the mesh. We plot both the real
simulation run-time data and the ideal ones for the number of cores we have in
each of the cases, and show the speedups for both. The ideal speedups are
calculated from the $t_r(N_{\mathrm{min}})$, as is the usual custom in such
scaling estimates. We can see that we obtain good strong scaling, as the
real speedups not only rise almost in a straight line,
but track the ideal lines
very closely. The fact that they overshoot somewhat
is because we have
only an estimate of $t_r(1)$ from $t_r(N_{\mathrm{min}})$.
This is the same reason why the efficiency curve in the lower
plot rises above $1$.
This confirms that \nmesh~scales efficiently with an increasing number of
cores or MPI processes.

\subsubsection{DG+FV/FD hybrid (S) simulation}
\label{stablehy}
\begin{figure}
	\includegraphics[width=\linewidth]{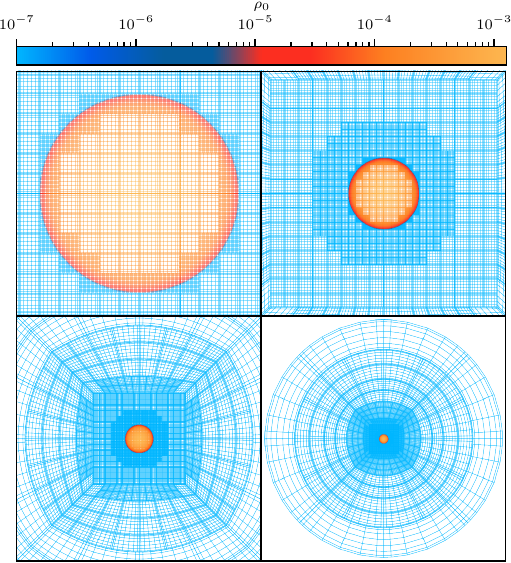}
	\caption{\label{stablhymesh} Initial mesh in the $xy$-plane
	from the (S) simulation with the hybrid DG+FV/FD evolution method.
	Top left: the star; top right: the central cube;
	bottom left: up to cubed sphere layer surrounding cube;
	bottom right: the entire mesh.
The elements with denser, equally spaced grid points, are the ones using FV/FD.
The DG elements are the ones with sparser, unequally spaced grid points. The
color map shows the rest mass density $\rho_0$ profile.
We see that only the troubled region
at the star surface is initially set to use FV/FD.
}
\end{figure}
We also evolve (S) with our hybrid DG+FV/FD scheme.
In the elements that use FV/FD,
we use the WENOZ\ts{2} reconstruction scheme as described
in Sec.~\ref{modFV}.
Before using
combined the DG+FV/FD method, we have
performed tests with pure FV/FD
with the WENOZ\ts{2} to validate our new method.
However, we chose case (U) for these instead of (S), as it is significantly
more challenging, and hence offers much more robust and general conclusions.
These results are presented in Secs.~\ref{fvmigpref} and \ref{mighref}.
For the WENOZ\ts{2} method in this case, we turn on extrapolation near the
boundary as per the $w = 4/3$ setup for Eq.~(\ref{fv_divf_extrap}) described
in Sec.~\ref{modFV}. Hence, we also switch on the cutoff levels for using $w=1$
as when $\rho_0 < 10^{-9} \rho_{0,max}(t=0)$ or
$\epsilon < 10^{-9} \epsilon_{max}(t=0)$, as described there. When the
surface points of neighboring elements do not coincide, we use
parabolic interpolation to exchange data between neighboring
elements at the boundary. In the FD scheme used for the GH system we use
fourth-order dissipation in the interior of the 
elements [see Eq.~(\ref{KO-diss})] with $\sigma = 0.1$.
Since we want the high-order WENOZ used in the interior of elements
to dominate the result, we increase the number of
points in each direction in each FV/FD element to $n = 16$, while in the DG
elements, we use half as many points ($n = 8$),
as DG can provide high accuracy
with lower number of points in the smooth regions. Each element then has
$n \times n \times n$ total points.
For this simulation, we set up the mesh similar to what we use for
case (B), which is described later in Sec.~\ref{boost}, as they are the
same star, with and without an imparted boost.
The detail of the initial mesh can be seen from Fig.~\ref{stablhymesh},
which shows plots on the $xy$-plane.
In the left top plot of the figure, we see the zoomed in view of the star.
Initially, we set FV/FD only at the troubled region at the
edge of the star. In order to determine which elements should be set to
FV/FD at the initial time, we look at the coefficients of $\rho_0$
in an expansion in Legendre polynomials, as described in Sec.~2.5
of~\cite{Tichy:2022hpa}.
As we go from lower to higher order coefficients, we find a much slower
dropoff for elements that contain the star surface, than for elements in the
star interior. We use this dropoff to initially set only the elements
containing the star surface to use FV/FD. These are recognizable
from the denser grid lines as FV/FD uses twice as many points as DG.
The initial number of DG and FV/FD elements across the star is
given in  Table \ref{NSres}.
Next, in the top right plot we see the
central cubical region, which is the best resolved region
of the mesh. Here the size of this cube is $62$.
However, as we can see, the resolution is not uniform. This is because
we do not apply uniform h-refinement to the entire mesh. We h-refine a 
spherical region of radius $15$ centered on the star to $l=5$ times,
which is the highest refined region in the mesh. The h-refinement is then
dropped by one level as we move away from the star, with the different
h-refinement spheres having radii $15, 30, 60,$ and $120$. These
successive regions can be seen
in the plots on the top right, bottom left, and bottom right. In the
left bottom picture, we see the mesh up to the first cubed sphere patches
surrounding the central cube, up to a distance of $104$.
The last bottom right picture shows the
entire mesh, which has a radius of $208$.
In total, we have 13 patches similar to the setup in Sec.~\ref{stableDG}.
In the most refined central region (containing the star) in this arrangement,
the elements have a dimension of
$1.625~(2399.516$ m). Of these, the DG elements in the interior
of the star have the following grid spacing:
$\Delta x_\mathrm{min,DG} = 0.0805~(118.813$ m),
$\Delta x_\mathrm{max,DG} = 0.362~(533.943$ m), and
$\Delta x_\mathrm{avg,DG} = 0.232~(342.788$ m).
On the other hand, the FV/FD elements on the star surface have
uniform grid spacing $\Delta x_\mathrm{FV/FD} = 0.116~(171.394$ m).
This is because we have twice the number of grid points in the FV/FD
elements compared to the DG elements.
The time step is set as per Sec.~\ref{time-int}.

For the atmosphere we use the
values $f_\mathrm{atmo,level} = 10^{-11}$, $f_\mathrm{atmo,cut} = 10^{-9}$,
and $f_\mathrm{atmo,max} = 1$,
as this is the setup we also use for (U) and (B) simulations with DG+FV/FD,
which are described later. The other atmosphere limits are still
maintained at $\epsilon_\mathrm{atmo,max} = 0.1$,
and $(Wv)_\mathrm{atmo,max} = 0.01$. Here, we
employ the entire limiting process described in Sec.~\ref{sec:lim} for $D$,
$\tau$, and $S_i$, including both elementwise and pointwise limiting.
For this purpose, we set $f_S^\mathrm{lim} = 0.99$.
In this simulation, we use the conversion procedure from conserved
to primitive variables stated in Sec.~\ref{cons2prim},
setting $\epsilon_\mathrm{floor} = 0.0$.
The constraint damping parameters for the GH system are set to be the
position dependent functions of the form from Eq.~(\ref{3gaussians})
along with Eqs.~(\ref{Par3Gf})-(\ref{Par3Gl}).
The gauge is chosen to be the harmonic gauge, where the
gauge source function is set as $H_\alpha = 0$.
In the DG elements, we filter the conserved matter
variables $D, \tau$, and $S_i$ with $s=40$, while using $s=64$
for the GH variables of the
evolved metric and its derivatives. We maintain
$\alpha_f = 36$ for both. No filters are used in the FV/FD elements, as
dissipation plays the counterpart role. When incorporating the
TEI described in Sec.~\ref{fvdg}, we set the
value of $ \ds \alpha_N^{\mathrm{evo}} $ from Eq.~(\ref{persson})
in the elements using DG to be $8$
and $9$ in the elements using FV/FD.
For this criterion,
we only consider the coefficients
that are greater than $f_\mathrm{unfiltered} = 0.9$ times their
unfiltered value. We set the cutoff from Eq.~(\ref{TroubleD}) as
$\rho_\mathrm{cut,low} = 10^{-4} \rho_{0,\mathrm{max}}(t = 0)$ and 
$\ds \rho_\mathrm{cut,hi} = \ds 10~ \rho_\mathrm{cut,low}$.
\begin{figure}
	\includegraphics[width=\linewidth]{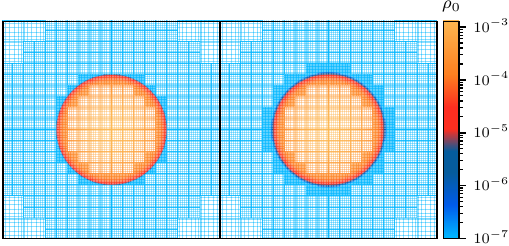}
	\caption{\label{stablerho} An $xy$-plane slice of the star
	from the (S) simulation with the hybrid DG+FV/FD method
at two times. Left: $t = 0 M_\odot$; right: $t = 1000 M_\odot$.
We see that only the star surface continues to evolve with FV/FD 
at $t = 1000 M_\odot$, showing proper TEI functioning. The TEI adapts
to the slight expansion of the star surface.}
\end{figure}

We show the density $\rho_0$ plotted as a colormap on the mesh wireframe
in the
$xy$-plane in Fig.~\ref{stablerho}, zooming in on the star. 
The initial setup can be seen in the left plot.
The right box of Fig.~\ref{stablerho}, shows $\rho_0$ after evolving
to time $t = 1000 M_\odot$.
We see that the star has expanded slightly from the initial configuration.
The number of elements using FV/FD also increases accordingly to adapt
to the star surface as it expands. Still, only the surface of the star
is evolved with FV/FD, indicating proper functioning of our TEI.

\begin{figure}
	\includegraphics[width=\linewidth]{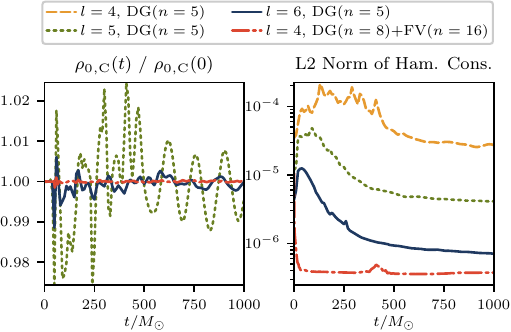}
	\caption{\label{stableDGhyComp} This figure compares results from DG+FV/FD 
	(S) simulation with pure DG (S) simulations for central density
	$\rho_{0,\mathrm{C}}$(left) and L2 norm of Hamiltonian constraint in the
	central cube (right). We see that the noise in $\rho_{0,\mathrm{C}}$ for the
	DG+FV/FD simulation is significantly lower than
	all the pure DG simulations. For the
	Hamiltonian constraint, the DG+FV/FD simulation settles down to its
	steady value faster than all DG simulations, and remains lower as well.
	}
\end{figure}
Lastly, we look at the central density $\rho_{0,\mathrm{C}}(t)$ 
normalized to the initial value (left plot) and the Hamiltonian constraint's
L2 norm in the central box (right plot) in Fig.~\ref{stableDGhyComp},
which are the representative quantities we study
for all the DG+FV/FD simulations, including for (U) and (B) described later.
For the Hamiltonian constraint, we compare all the
(S) simulations. For $\rho_{0,\mathrm{C}}(t)$, we exclude the lowest
resolution pure DG run as we know it has
large noise in $\rho_{0,\mathrm{C}}(t)$.
It should be kept in
mind when making these comparisons that the bulk of the mesh for our DG+FV/FD
scheme has a resolution that falls in between the $l = 4$ and $l = 5$
cases for the pure DG simulations in Sec.~\ref{stableDG}. Even though the
central h-refinement sphere for DG+FV/FD 
uses $l = 5$, the dimension of the central
cube is much larger than the previous setup.
However, as the DG scheme is expected
to demonstrate exponential convergence when increasing the order
of the basis polynomials, the choice of $n = 8$
for the DG elements in this DG+FV/FD
evolution should give us much more accurate results for the same h-refinement
level. As was mentioned in Sec.~\ref{stableDG}, the $l = 4$ level was kept in
the discussion only as a reference for improvement of results with increasing
resolution. We started getting reliable
results only from the $l = 5$ resolution
onwards. Keeping all these in mind, we expect the DG+FV/FD case to be
the hybrid counterpart closest to the pure DG scheme of $l = 5$ resolution.
As can be seen from the left plot for $\rho_{0,\mathrm{C}}(t)$ in
Fig.~\ref{stableDGhyComp}, the central density
for DG+FV/FD essential settles down immediately
into systematic oscillations compared to the pure DG runs.
Additionally, the oscillation amplitude is much smaller.
This is because the FV/FD at the surface of the star
eliminates the Gibbs oscillations that
plague the pure DG simulations at early times, when the
perturbations from the numerical errors kick in.
This better behavior of the central density oscillations are
also reflected in the corresponding spectrum in Fig.~\ref{dgDfft}.
We see that the first three peaks for the (S) DG+FV/FD case are
much more prominent than for the pure DG (S) simulation.

This improvement is also reflected
in the Hamiltonian constraint shown on the right of
Fig.~\ref{stableDGhyComp},
which settles down much faster for the DG+FV/FD case.
In fact, the results settle to a steady state
even faster than the $l=6$ pure DG run,
and to an even lower value.
The comparable case of $l=5$ still remains
at a level much above the DG+FV/FD case.
Hence, we can see that the DG+FV/FD simulation performs significantly better than
the equivalent pure DG simulation.

\subsection{Perturbed configuration (P)}
\label{perturb}

Having confirmed the robustness of our program through (S),
we next apply a pressure perturbation to the star at $t=0$.
We call this case (P).
The pressure perturbation is
of the form $P + \delta P$~\cite{Baiotti:2008nf}, with
\be
\label{pertP}
\delta P = \ds \lambda\cdot (P + \rho_0 + \rho_0\epsilon)
						\sin\left(\ds \frac{\pi r}{r_\mathrm{surf}}\right)
           Y^0_2(\theta,\varphi),
\ee
where $(r,\theta,\varphi)$ are the standard spherical
coordinates, $Y^0_2(\theta,\varphi)$ is the $l=2$, $m=0$
spherical harmonic, and $r_\mathrm{surf}=8.1251439$ is the radius of the
unperturbed star. Both $r$ and $r_\mathrm{surf}$ are in isotropic coordinates,
which is the coordinate system used in the cubical patch in
the center of our mesh in which we evolve the star.
Once $P$ is set, we recompute the 
corresponding values for $\rho_0$ and $\epsilon$
for the initial setup using the
EoS in Eq.~(\ref{inidataeos}). The metric variables are left unchanged
and maintain their unperturbed TOV values as in (S). The magnitude of the
perturbation parameter is set to $\lambda = 0.05$, which is a reasonably strong
perturbation.

\begin{figure}
	\includegraphics[width=\linewidth]{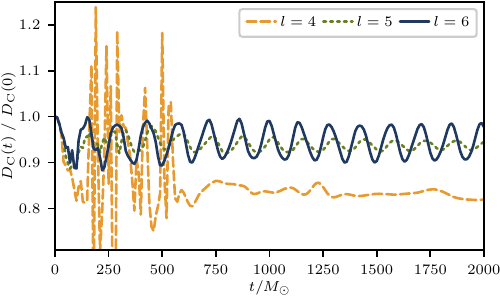}
	\caption{\label{dgPerturbedDpt0}
  This figure shows central density
	$D_\mathrm{C}$ for (P) with pure DG simulations corresponding to
	h-refinement levels of $l = 4, 5$ and $6$, normalized to the initial
	value. Same as for (S), substantial Gibbs oscillations 
	spread across the star from the surface, generating large amounts of
	noise, especially for the low resolution.
  Once again, as resolution increases, systematic oscillations consistent
	with principal frequency of the star becomes more pronounced.}
\end{figure}

The remaining setup, apart from the pressure perturbation, remains exactly
the same as the pure DG (S) simulation in Sec.~\ref{stableDG}. To study the effect
of different resolutions, we again look at the
central conserved density $D_C$ plot for the different refinement levels
of $l=4, 5$, and $6$, shown in Fig.~\ref{dgPerturbedDpt0}. In this case,
as there is an actual relatively stronger perturbation present, the oscillations
in the central density are much more pronounced than in (S) where the
oscillations were simply the result of much weaker perturbations from numerical
noise. Again, we see that the $l=6$ case displays the expected behavior the
best, while the lowest resolution of $l=4$ is once again swamped by Gibbs
oscillations and numerical noise to the extent that we are unable to observe
the characteristic oscillation we expect in this case.
We again look at the Fourier transform of the central
density in Fig.~(\ref{dgDfft}) for the $l = 6$ h-refinement case.
The finest element dimension and grid spacing detail for this
simulation is the same as mentioned for the corresponding $l=6$ pure
DG (S) case in Sec.~\ref{stableDG}. Although the perturbation is significant,
the deviation in the physical attributes such as mass and central density
are within a margin such that characteristic spectrum is still in agreement
with the (S) configuration as described in \cite{Font:2001ew}. Hence, we
still see the principal mode to be in perfect agreement with that predicted by
theory. In fact, it is more pronounced as the oscillation is larger
than (S). Once again, the higher modes are swamped by numerical errors.

The takeaway is that \nmesh~is able to successfully evolve a perturbed star
with gravity coupled to the matter fields with pure DG.
The results agree with literature 
results, as can be seen from the principal oscillation frequency from the
Fourier spectrum.

\subsection{Unstable configuration (U)}

Having tested that our program successfully and stably handles the dynamics
of the perturbed star in case (P), we move on to the case
of a star with an initial configuration of a TOV
star on the unstable branch (U), where
$\p m_\mathrm{ADM} / \p \rho_{0,\mathrm{C}} < 0$~\cite{CorderoCarrion:2008nf,
Carvalho:2019ert}. Here $m_\mathrm{ADM}$ is the
so-called ``ADM'' mass of the star being defined from the
Arnowitt-Deser-Misner formalism~\cite{Arnowitt62}.
The initial data for this star is in a highly
dense unstable equilibrium state~\cite{Font:2001ew}.
As such, any initial configuration on the unstable branch migrates
either towards a TOV configuration on the stable branch
or collapses into a black hole depending on the
nature of initial perturbations~\cite{Gourgoulhon1992a,Baiotti:2003btu,Font:2001ew,
CorderoCarrion:2008nf,Radice:2010rw,Radice:2011qr,Carvalho:2019ert}.
In this article, we want to study the case where the initial unstable 
TOV configuration migrates towards a
stable TOV configuration. It is thus important to have an initial
perturbation that drives the system in this direction.

This case is significantly more challenging than the previous ones, as the
matter fields are much more dynamic and there are shocks, which are
most visible in the $\epsilon$ field~\cite{Font:2001ew,Thierfelder:2011yi}.
Even though the matter fields such as $\rho_0$ and
$\epsilon$ were not smooth at the star surface in (S) and (P), the nonsmooth
feature remained relatively localized to its initial position as the
configuration did not change drastically. The only motion came from the
comparatively much gentler oscillations arising from either numerical or
the pressure perturbation.
However, the star (U) expands and contracts violently as it gets
displaced from the initial configuration. It shoots past its stable
configuration and goes back and forth between relatively higher and lower
central density configurations. For handling these violent motions and shocks
in the matter fields, the DG method alone was not suitable.
Even though the system did evolve without crashing with DG alone,
the dissipation caused by the filters and limiters was so significant,
that the results, at affordable resolutions, were not close to
the well-known expected behavior as 
stated in~\cite{Font:2001ew,Baiotti:2008nf}. Hence, to handle this case
we had to fall back to the FV/FD method, which provides robust
HRSC capabilities.

As our FV/FD method is a modified one, we started out by first evolving
the case with pure FV/FD with WENOZ\ts{2}.
We ran simulations with different resolutions
for (U) to test behavior of results on the ANVIL supercomputer.
We tested the results separately for p-refinmenet and h-refinement variations.
We still used the mesh setup of a central cube surrounded by cubed sphere
patches, with total $13$ patches as for (U) and (P). However, as (U) is
initially much denser, it has a much smaller radius, and to better adapt
to this smaller size, we set the length of the central cube to $16$.
We use the position dependent constraint damping from
Eq.~(\ref{3gaussians}) and Eqs.~(\ref{Par3Gf})-(\ref{Par3Gl})
and harmonic gauge with $H_\alpha = 0$
for the GH system.
The atmosphere is set such that $f_\mathrm{atmo,level} = 10^{-11}$ and
$f_\mathrm{atmo,cut} = 10^{-9}$, $f_\mathrm{atmo,max} = 1$,
$\epsilon_\mathrm{atmo,max} = 0.1$, and $(Wv)_\mathrm{atmo,max} = 0.01$
as introduced in Sec.~\ref{atmo}.

\subsubsection{Varying p-refinement for pure FV/FD (U) simulations}
\label{fvmigpref}

\begin{figure*}
  \includegraphics[width=\linewidth]{
    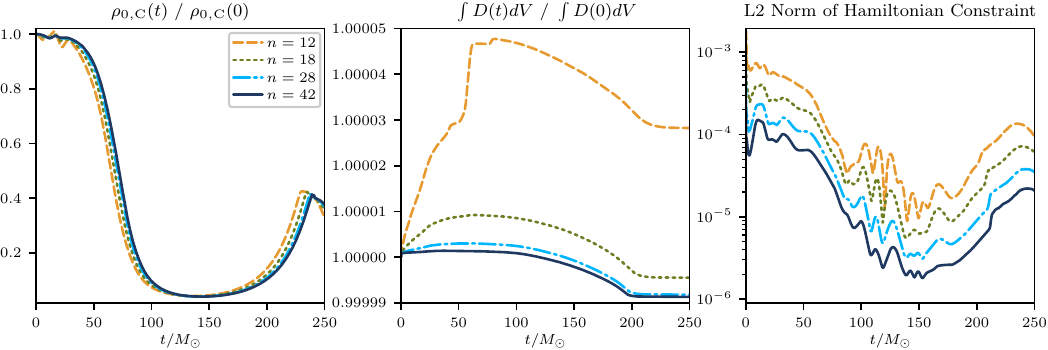}
  \caption{\label{fvmigpreftData} Results for p-refinements of
  $n = 12, 18, 28$, and $42$ and h-refinememnt level of $l = 3$
  for the pure FV/FD simulation of (U) till the first
  peak when the star contracts.
  Left: central rest mass density
  $\rho_{0,\mathrm{C}}$(normalized to the initial value);
  center: rest mass $\int D dV$ in the central cube
  (normalized to the initial value);
  right: L2 norm of Hamiltonian constraint in the  central cube.
  We see that the peak value of $\rho_{0,\mathrm{C}}$ is in good
  agreement with that in other simulations such as in~\cite{Font:2001ew}.
  The offset in time
  of the different resolutions is because the star gets displaced from the 
  initial state at different times due to difference in perturbations from 
  the different resolutions. The rest mass and the Hamiltonian
  constraint show convergent behavior.}
\end{figure*}

For the case of p-refinement, we vary the number of points in each element
in each direction to be $n = 12, 18, 28$, or $42$, with five elements
across the star.
Since we increase
the number of points from the pure DG (S) and (P) setups and also
made the central cube smaller, we lower
the h-refinement level for these runs to $l=3$, so that the elements
are larger, resulting in $6656$ total elements. The parameters for the
limiters (see Sec.~\ref{sec:lim}) are the same as in Sec.~\ref{stablehy}.
When converting from conserved to primitive variables, the procedure
in Sec.~\ref{cons2prim} is relaxed, and for this case we allow
negative $\epsilon$ values by setting $\epsilon_\mathrm{floor}=-0.1$.
Because of this, $\epsilon$ dips below $0$  for short periods of time
in places where a shock forms when matter falls back onto the star during
contraction
due to numerical errors in the lower resolutions $n = 12$  and $n = 18$.
Yet, for the higher resolutions of $n = 28$ and $n = 42$, $\epsilon$
remains positive.
We do not apply filters to any of the variables here, as we
are evolving with pure FV/FD and our WENOZ\ts{2} scheme is capable of handling
any nonsmooth features. However, here we set
$w = 1$ in Eq.~(\ref{fv_divf_extrap}), thus dropping extrapolation
near the boundary. At the boundary, when interpolation is needed for data
exchange between neighbors, we use linear interpolation. One point
away from element boundaries, we also add a second-order dissipation
operator to the GH system, by using the standard centered
second-derivative stencil there.
The dissipation factor is set to $\sigma = 0.1$. 
These extra modifications were done out of an abundance of caution. However,
later studies have shown that they are not needed to keep the system stable.
Consequently, they were not used in the more sophisticated hybrid simulations
discussed below.

For the simulations with varying
p-refinement discussed in this section and those with varying h-refinement
discussed in the following Sec.~\ref{mighref}, there was a discrepancy
in our implementation of the derivatives $\p_i \alpha$ needed in the
source terms for the hydrodynamic
variables outlines in Eq.~(\ref{GRHDsources}). This was later discovered
and fixed for all the other simulations. This discrepancy produced a
maximum of $1.5\sim2\%$ difference in the source terms,
depending on the term considered. Incidentally, this maximum occurs
very early in the simulation, just when the star gets displaced from
its initial unstable equilibrium configuration. The magnitude of this
difference then decreases significantly during the rest of the
simulations. As such, this discrepancy essentially functions like
a perturbation, which coincidentally has a form that drives the star
towards the stable configuration. The magnitude of this perturbation is
then also comparable to the perturbations usually applied in such
studies~\cite{Radice:2011qr,Baiotti:2003btu} and also to the 
density perturbation we use in Sec.~\ref{mighy}.

Although the lower resolutions were continued
until about a time of $t=900 M_\odot$,
we look at the results until time $t=250 M_\odot$, where we have stopped
our higher resolution runs because of computational cost. As can be seen 
from the left plot in Fig.~(\ref{fvmigpreftData}), which shows the
variation of central rest mass density $\rho_{0,\mathrm{C}}$,
we have the simulation
continuing through the first expansion phase of the star, starting from
the initial highly dense unstable state, until it partially contracts back
again for the first time.
The oscillations seen in the beginning result from perturbations
due to numerical errors, while the star is still hovering
around the unstable equilibrium before migrating towards the stable one.
As the star gets displaced from its initial state partially by perturbations
arising from numerical errors, the time at which the star starts to expand from
the initial compact state differs for the different resolutions. As such,
the first major expansion phase is offset by different amounts in time for
different resolutions.

For the reasons discussed below,
we shift the star away from the center of the mesh
for these (U) simulations.
Due to h-refinement, the corners of eight elements meet at the mesh
center. If the star is centered on the mesh, sharp features in the density
form at the center in our simulations.
Similar sharp features have been observed
for this unstable star system in~\cite{CorderoCarrion:2008nf,Radice:2011qr}.
However, simulations with conventional FV schemes as in~\cite{Font:2001ew}
do not show such sharp features. As explained in~\cite{Radice:2011qr},
we believe that these features cannot be easily resolved by conventional
FV methods, and are therefore not shown in much of the literature. 
Hence, to obtain results as close
as possible to the standard 2D or 3D simulations for this
setup~\cite{Font:2001ew,Baiotti04a,Thierfelder:2011yi},
and
for ease of comparison and vetting of results,
we have decided to move the mesh so that the
star center [located at the $(0,0,0)$] is at the center of an element.
This suppresses sharp features in our simulations, as our FV/FD
scheme acts like any standard FV method at element centers.
For the simulations discussed below, the mesh is centered
on the point $(1,1,1)$.
Even with this setup there still remain some oscillations in the
central density $\rho_{0,\mathrm{C}}$.
However, the overall nature of our results agrees
with standard results obtained with FV methods in three dimensions.
Additionally, the variation of $\rho_{0,\mathrm{C}}$ with resolution
in left plot of Fig.~(\ref{fvmigpreftData})
is consistent with second-order convergence.

From the center plot in Fig.~\ref{fvmigpreftData} we see that
mass conservation improves with resolution.
For the highest resolution case, the mass remains almost constant until about
a time of $t=100 M_\odot$. Until this point,
the bulk of the matter is contained in the
innermost cube, which is the best resolved region. However, since the star
is expanding, the matter gradually spills over into the surrounding cubed
sphere patches. Since linear interpolation takes place at the element boundary
between an element in the central cubical patch and an adjacent element in
the cubed sphere region, the mass does not remain exactly conserved, causing
the drop we see around this time. Furthermore, the resolution also drops in
the elements in the cubed sphere region, which also contributes to the
mass loss we observe. Lastly, the limiters and atmosphere treatment we apply
perform pointwise fixes, that also contribute to this violation. At around
$t=160 M_\odot$ the central higher density region, containing matter with 
density above $10^{-2}  \rho_{0,\mathrm{C}}(t = 0)$, stops expanding,
and starts contracting again. However, a lower density shell of about
$10^{-2} \rho_{0,\mathrm{C}}(t = 0)$ to  $10^{-5}
\rho_{0,\mathrm{C}}(t = 0)$ continues to
expand, going further into the less resolved region, which causes the
$\int D dV$ quantity to keep dropping. This shell stops expanding at around
$t = 192 M_\odot$, which is when we see
the second stabilization of the mass.
The overall takeaway again is that the mass conservation improves
drastically with increasing resolution.
The error caused by the interpolation, and the
other factors converge away with increasing resolution, providing reliable
results as we go to reasonably good resolutions.

Lastly, we look at the L2 norm of
the Hamiltonian constraint violation of the
central cube as although some matter eventually flows out of it, the
majority of the matter still remains contained within it.
Again we see convergent behavior with increasing resolution.
For the constraint violations shown in Fig.~\ref{fvmigpreftData}, we find a
convergence order of $1.5$.

Overall, we see that our results for the simulations agree with the expected
behavior of the system as per standard literature. Additionally, results
improve with resolution in systematic fashion for all the quantities we study.


\subsubsection{Varying h-refinement for pure FV/FD (U) simulations}
\label{mighref}

The setup of the patches in the mesh remains the
same here as in Sec.~\ref{fvmigpref}.
However, we use $n=16$ points in each direction in each of our elements.
Furthermore, we no longer apply uniform h-refinement, and instead
switch to the concentric  spherical decreasing
h-refinement regions centered on the star (see Fig.~\ref{stablhymesh}),
introduced in Sec.~\ref{stablehy}.
\begin{figure*}
	\includegraphics[width=\linewidth]{
		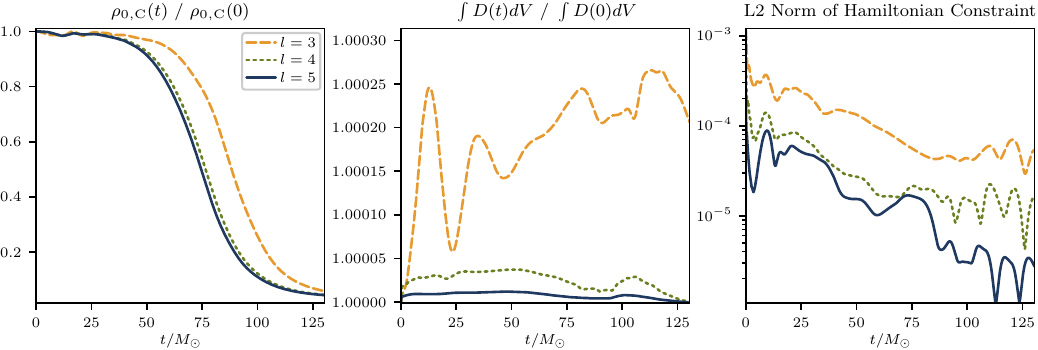}
	\caption{\label{fvmighreftData} Results for
	central h-refinements of $l = 3, 4$, and $5$ with $n = 16$
	for the pure FV/FD simulation of (U). Left: rest mass density
	$\rho_{0,\mathrm{C}}$ (normalized to the initial value);
  center: rest mass in central cube (normalized to the initial value);
  right: L2 norm of Hamiltonian constraint in central cube.  }
\end{figure*}
The radius of the most highly refined region
is chosen to be $30.5$ to ensure that the bulk of the matter is contained
inside this region. The refinement is dropped by one level after radii
$30.5, 61, 122$ from the origin, with the radius of the entire mesh
being $200$. The resultant initial number of elements
across the star for this 
setup are stated in Table \ref{NSres}.
The central region refinements are chosen to
be $l = 3, 4, 5$ for the different cases.
The center of the mesh is shifted to $(1,1,1)$,
$(0.5,0.5,0.5)$, and $(0.25,0.25,0.25)$ for the three different resolutions
respectively, to always have the star center at the middle of an element,
similar to Sec.~\ref{fvmigpref}.
The FV/FD related setup for this run is the same
as the treatment of the FV/FD elements in the
DG+FV/FD run for (S) in Sec.~\ref{stablehy}.
This means we switch on the extrapolation near the element
boundary for WENOZ\ts{2},
use parabolic interpolation at the boundary, and
limit dissipation to the interior of an element as in 
Eq.~(\ref{diss-matrix}).
The hydro setup also remains
the same as in Sec.~\ref{stablehy}, and so does the GH system.

\begin{figure}
	\includegraphics[width=\linewidth]{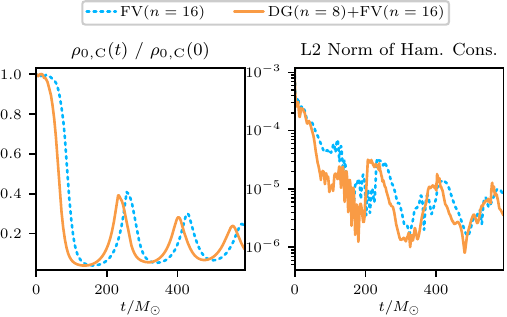}
	\caption{\label{migFVhycomp} This figure shows $\rho_{0,\mathrm{C}}$
	normalized to the initial value (left) and
	L2 norm of Hamiltonian
	constraint in central box (right) for the comparable pure FV/FD and DG+FV/FD
	cases for (U).
	Both have central h-refinement of $l = 3$ and use $n = 16$ in their
	FV/FD elements.
        Both simulations are in	good agreement for both quantities.
        Furthermore, $\rho_{0,\mathrm{C}}$ agrees well
	with standard literature values~\cite{Font:2001ew}.}
\end{figure}

The behavior of the different quantities such as $\rho_{0,\mathrm{C}}$,
$\int D dV$, and L2 norm
of the Hamiltonian constraint in the central cube
remain similar to that for varying p-refinement,
as can be seen from the left, center, and right plots
in Fig.~\ref{fvmighreftData}
respectively. We see systematic improvements in all the quantities with
increasing resolution. Unfortunately, due to high computational cost, we could only
continue the highest resolution simulation to barely beyond the
first expansion phase. This limits our capability to
make strong conclusions, but we can still see that the
results improve with resolution and are trending in a desirable fashion.
The lowest resolution simulation was continued for much longer, the results for
which can be seen in Fig.~\ref{migFVhycomp} by looking at the dashed line
curves.
We see the characteristic alternate
peaks and troughs in the central density $\rho_{0,\mathrm{C}}$ with
attenuating amplitude in the left plot,
that is known to be the appropriate behavior of (U) setup when evolving
with an EoS that allows for heating.
In the right plot we see that the L2 norm of the Hamiltonian constraint also
goes through the undulations as the matter flows in and out of the central box,
while maintaining overall stability. Given that the lowest resolution shows
stable and accurate behavior in the long term simulation, and the shorter
runs show convergent behavior with increasing resolution for the various
quantities, we can cautiously conclude that this setup for the simulations
are handled well with~\nmesh, and we expect that we will be able
to see proper convergent behavior with
longer runs and more different resolutions
in our future work.
Having said that, for these short duration runs, we find that
$\rho_{0,\mathrm{C}}$ in Fig.~\ref{fvmighreftData}
is consistent with convergence of order $2.55$ with resolution,
while the Hamiltonian constraint
L2 norm in Fig.~\ref{fvmighreftData} tends towards second-order convergence.


\subsubsection{DG+FV/FD (U) simulation}
\label{mighy}

Having confirmed the robustness and soundness of our new FV/FD method, we
next evolved (U) with the hybrid DG+FV/FD method. For this case,
the setup related to the mesh, WENOZ\ts{2}, FV, FD, metric and
hydro evolutions remain same as cases with h-refinement
$l = 3$ and $4$ and number of points $n = 16$ in the previous
Sec.~\ref{mighref}. However, we now incorporate our TEI
exactly the same way we did for the DG+FV/FD simulation of (S)
in Sec.~\ref{stablehy}.
The associated treatments of all the aspects of the DG elements also remain the
same as in Sec.~\ref{stablehy}. 
For these simulations, we apply a perturbation to $\rho_0$
of the form
\begin{equation}
\label{pertrho0}
\delta \rho_0 = \ds \frac{\lambda \rho_0}{\sqrt{4 \pi}} 
									\cos \left( \frac{\pi r}{2 r_\mathrm{surf}} \right)
\end{equation}
at the initial time.
Here $r=\sqrt{x^2+y^2+z^2}$ is the distance from the star center in
isotropic coordinates, and $r_\mathrm{surf}=4.2677044$ is the radius of the
unperturbed star (also in isotropic coordinates). We choose $\lambda = -0.01$
so that the magnitude of the perturbation remains within $1\%$ as is the
usual order of magnitude chosen for such
perturbations~\cite{Radice:2011qr,Baiotti:2003btu}.
Additionally, the sign is chosen to drive the star towards the stable
TOV configuration instead of a black hole. The reason for choosing this
perturbation is that if left solely to perturbations arising from numerical
errors, the star would sometimes collapse to a black hole, depending on the
resolution, mesh setup and other such factors. Since here we wish to study
the migration of an unstable star to a stable TOV star, we use this
perturbation to ensure that the unstable star is consistently pushed in this
direction, and not in the direction of forming a black 
hole~\cite{Gourgoulhon1992a,Baiotti:2003btu,Font:2001ew,
CorderoCarrion:2008nf,Radice:2010rw,Radice:2011qr,Carvalho:2019ert}.

\begin{figure}
	\includegraphics[width=\linewidth]{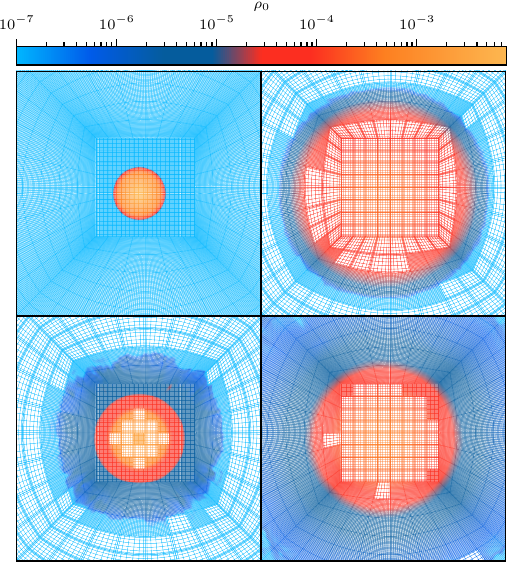}
	\caption{\label{migrho} Part of the mesh
	in the $xy$-plane from (U)
simulation with hybrid DG+FV/FD evolution
method at four different times. Top left: $t = 0 M_\odot$ (initial state);
top right: $t = 136 M_\odot$ (first minimum of
$\rho_{0,\mathrm{C}}$); bottom left: $t = 232 M_\odot$
(next peak with $\rho_{0,\mathrm{C}} \sim 0.4
\rho_{0,\mathrm{C}}(t = 0)$);
and bottom right $t = 320 M_\odot$ (second minimum
of $\rho_{0,\mathrm{C}}$).
FV/FD elements have denser equally spaced points,
while DG has sparser unequally spaced points.
We see that our hybrid scheme's TEI successfully tracks
the low-density (shown through the colormap)
troubled region as the star expands and contracts.
}
\end{figure}

We again check that our TEI is working properly
by verifying that only the elements near nonsmooth regions switch to
FV/FD. Figure~\ref{migrho} shows
the density on the mesh wireframe, similar to
Sec.~\ref{stablehy}, to check
which parts have switched to FV/FD, while zooming in on the central cube
in the $xy$-plane.
Unlike for (S), we cannot focus only on the central cube this time, as the
matter flows out into the rest of the mesh over time.
We pick four time frames to look at in this figure. The top left figure shows
initial $t = 0 M_\odot$ setup, where we set
every element to FV/FD. This is because
the star surface has a much steeper rise in the matter fields, and so we simply
set all elements to FV/FD initially to be cautious, letting
the TEI set the appropriate elements to DG after 11 time steps.
The next plot (top right) at $t = 136 M_\odot$
is close to when the central density of the star reaches the first minimum
after the star expands, in the context of the left plot of Fig.~\ref{migFVhycomp}.
Only a narrow troubled region at the boundary of the star
uses FV/FD, which has
matter density between $10^{-2} \rho_{0,\mathrm{C}}(t = 0)$ and
$10^{-4} \rho_{0,\mathrm{C}}(t = 0)$.
The rest of the mesh elements are evolved with DG.
The next (bottom left) time shown is $t = 232 M_\odot$, which is close to the time
when the star first fully recontracts, producing the central maximum of
$\rho_{0,\mathrm{C}}(t) \sim 0.4 \rho_{0,\mathrm{C}}(t = 0)$ in the left plot of
Fig.~\ref{migFVhycomp}. The shell of FV/FD elements also follows the star surface.
However, now the star is starting to get larger, with the density range
$10^{-2} \rho_{0,\mathrm{C}}(t = 0)$ to $10^{-4} \rho_{0,\mathrm{C}}(t = 0)$
spreading further from the bulk of the matter in the center.
It should be kept in mind that even though we are looking only at the
$\rho_0$ field here, there are sharp features, and even shocks that
are particularly visible in the
$\epsilon$ field, linked to these fast-moving low-density shells, which
has also been observed in~\cite{Font:2001ew,Thierfelder:2011yi}.
The TEI picks up on them and switches from
DG to FV/FD as necessary.
Hence, a thicker
band of elements around the star switches to FV/FD. The center has also adopted
FV/FD as there are oscillations at the center due to matter falling
into it from all sides when the star contracts.
Similar oscillations have been
also observed in~\cite{Radice:2011qr}, when dissipation is low enough
compared to more traditional schemes. Lastly, we show time $t = 320 M_\odot$ in
the bottom right plot, corresponding to the second minimum in the left plot
of Fig.~\ref{migFVhycomp} when the star has once again expanded. Here we see
that the
troubled region has spread considerably and most of the field of
view in the plot is covered by FV/FD elements, while DG is used in the
smooth interior of the star. Of course, if we zoom out,
most of the rest of the
mesh is still largely evolved with DG. From these plots, we can determine that
our TEI functions properly, and can track the troubled
region appropriately even in a case like (U) where the star is undergoing
much more violent motions through expansion and contraction than the
considerably benign case (S) of Sec.~\ref{stablehy}.

\begin{figure}
	\includegraphics[width=\linewidth]{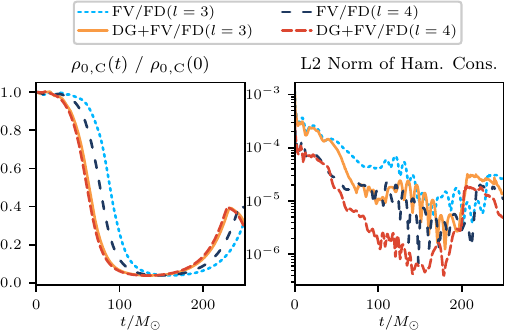}
	\caption{\label{migFVhylcomp} This plot shows
	central density $\rho_{0,\mathrm{C}}$ normalized to the initial value (left)
	and L2 norm of Hamiltonian
	constraint in the central cube (right) for comparable cases with pure 
        FV/FD and DG+FV/FD for (U) with central h-refinement levels of $l = 3$
	and $l= 4$. We see that for $\rho_{0,\mathrm{C}}$ the corresponding runs
	with pure FV/FD and DG+FV/FD are in good
	agreement. For the constraint violation,
	DG+FV/FD performs somewhat better than the
	pure FV/FD runs in the central cube,
	both in terms of the overall violation level, and the convergence behavior.}
\end{figure}

We also check the behavior of $\rho_{0,\mathrm{C}}(t)$ and L2 norm of
Hamiltonian constraint in the central cube for the DG+FV/FD runs with two
different h-refinement levels,
and compare them to the same quantities
from our FV/FD runs with corresponding
resolutions, as shown in Figs.~\ref{migFVhycomp} and
\ref{migFVhylcomp}.
The lowest resolution runs provide comparison
for longer time, as they are computationally cheapest.
We see that the results for $\rho_{0,\mathrm{C}}(t)$
with DG+FV/FD runs match reasonably well with their same
resolution pure FV/FD counterparts, considering the time
offsets owing to difference in nature of the initial perturbations.
The Hamiltonian constraint is lower for the same resolution
for the DG+FV/FD simulation than the pure FV/FD simulation.
Additionally, the Hamiltonian constraint improves from $l=3$ to 
$l=4$ resolution when considering only the DG+FV/FD curves,
while $\rho_{0,\mathrm{C}}(t)$ shows similar behavior.
These give us confirmation that
the DG+FV/FD evolutions are performing well.

\begin{figure}
	\includegraphics[width=\linewidth]{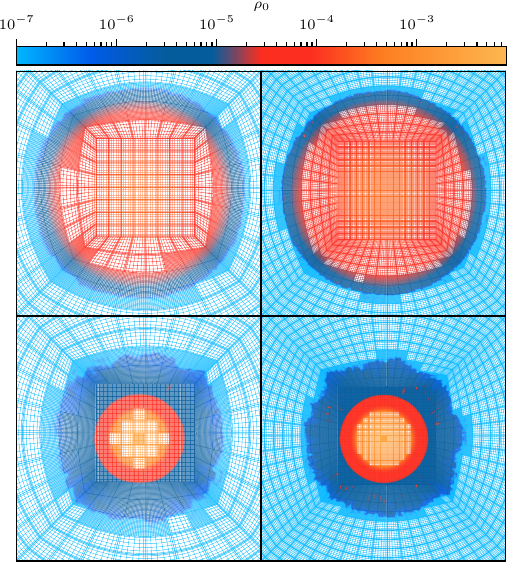}
	\caption{\label{migrhol} A portion of the
	mesh in the $xy$-plane from the (U)
simulation with the hybrid DG+FV/FD
method with different central h-refinmenet levels.
Top left: $t = 132 M_\odot$ for $l=3$; top right: $t = 148
M_\odot$ for $l=4$; bottom left: $t = 232 M_\odot$ for $l=3$; and
bottom right: $t = 228 M_\odot$ for $l=4$. Both show $\rho_{0}$ close to the
time when $\rho_{0,\mathrm{C}}$ reaches its first
minimum and the following maximum.
We see that the width of the region evolved with FV/FD decreases with
increasing h-refinement, as the troubled region is 
concentrated in the nonsmooth region near the star surface.
}
\end{figure}

Last, we compare the TEI performance for the DG+FV/FD simulations
with different levels of h-refinement
in Fig.~\ref{migrhol}.
This provides another crucial test, 
as with increasing h-refinement the TEI should
hone in on the troubled regions near the star surface,
thus using FV/FD in a smaller region. 
We see this successfully carried
out in Fig.~\ref{migrhol}, where the lower h-refined simulation in the
left plots show a thicker region of FV/FD elements,
as compared to the counterparts from the higher
h-refined simulation on the right.
The top plots correspond to the time when the star has expanded
fully for the first time, and $\rho_{0,\mathrm{C}}$ has
reached its first minimum, while the bottom plots are for the following
maximum in $\rho_{0,\mathrm{C}}$, when matter has fallen in.
For both these states of the star, the times for the two resolutions
of $l=3$ and $l=4$ have been chosen while accounting for the
time offset in the dynamics of the two resolutions to achieve
a faithful comparison. So, we can conclude from these plots that the
TEI performs properly and retains DG in more elements as the
element size decreases with increasing h-refinement.

Hence, from all these results, we can conclude that our TEI
can successfully handle not only the simple case of (S), but also the far
more challenging case of (U). Additionally, the results obtained from the
combined DG+FV/FD simulations agree well
with the expected results for this scenario,
and also improve with increasing resolution.

\subsection{Boosted configuration (B)}
\label{boost}

\begin{figure}
	\includegraphics[width=\linewidth]{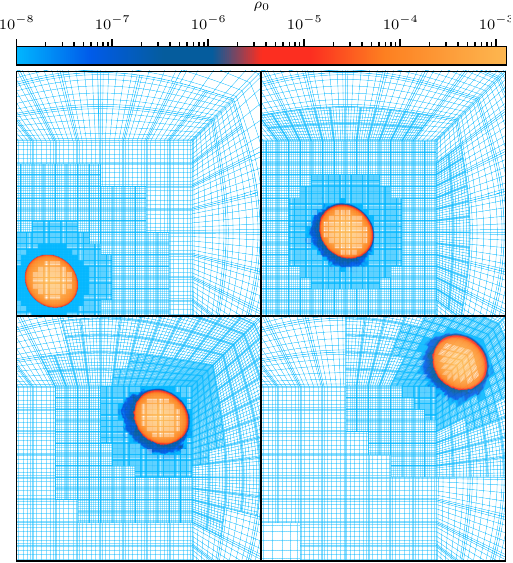}
	\caption{\label{boostrho} A portion of the
	mesh in the $xy$-plane from the (B)
simulation with $v_\mathrm{CoM} = 0.5$ using the DG+FV/FD evolution
method at four different times. Top left: $t = 0 M_\odot$; top right: $t = 40
M_\odot$; bottom left: $t = 88 M_\odot$; and bottom right $t = 132 M_\odot$.
The elements with denser, equally spaced grid points, are the ones using FV/FD.
The DG elements are the ones with sparser, unequally spaced grid points. The
colormap shows the rest mass density $\rho_0$.
We see that the TEI
nicely tracks the surface of the star and uses FV/FD only in that region.
Furthermore, we also see the adaptive mesh refinement spheres track the star.
The simulation continues smoothly even beyond the central cubical region.}
\end{figure}

The last case we study is the boosted star (B), that has an initial boost
applied to (S). We evolve two (B) setups with boost applied such that
the CoM of the star has a velocity
of $v_\mathrm{CoM} = 0.1$ and $0.5$ diagonally along the
$xy$-plane. (B) again proves too difficult to evolve with pure DG,
as large unphysical Gibbs oscillations develop that reduce the accuracy
significantly. Hence, we again turn to the DG+FV/FD method.
Almost all aspects of the simulation setups for (B) remain the same
as the (S) simulation in Sec.~\ref{stablehy},
with a few modifications, as  we impart a boost to
that very star in the simulations of this section.
The larger dimension of $62$ of the cube, compared to the (U) setup
and pure DG (S) setup, is chosen for allowing the star to travel
longer in the  well-refined cubical patch.
The resultant initial number of elements both parallel and perpendicular
to the $\vec{v}_\mathrm{CoM}$ direction are stated in Table \ref{NSres}.
This differs for the $0.5$ case for the two directions, but remains the
same for $0.1$.
The central h-refinement region is refined $l =  5$ times
for the $v_\mathrm{CoM} = 0.5$ case
and  $l = 4$ and $l =  5$ times for the $v_\mathrm{CoM} = 0.1$ case
we simulate, and the h-refinement falls off
successively, the same as in Sec.~\ref{stablehy}.
This higher level of h-refinement is again set to account for the large
dimension of the central box and maintains a
comparable number points as the other simulations with DG+FV/FD.
Unlike the situations for (S), (P) and (U), where the CoM of
the star does not move from its initial position, for (B) we track the star
center by following the minimum of the lapse, and use 
AMR to make the whole h-refinement arrangement
follow the moving star.
For efficiency, the mesh is not updated after every time step,
but rather after
regular short intervals of $\delta t = 0.7071067811865475$.
We also use load balancing and redistribute the elements among
all available MPI processes every time the mesh is updated.
The balancing is based on the measured time each element spends on doing 
calculations to determine the workload of each element, with the goal
of giving each MPI process the same amount of work.
All the other setups remain same as that
in Sec.~\ref{stablehy}  in all aspects.

We mainly concentrate here on the efficiency of our TEI.
In the wireframe colormap plot of $\rho_0$ in the $xy$-plane
for $v_\mathrm{CoM} = 0.5$ in Fig.~\ref{boostrho},
the left top plot is the initial setup. Here, we have
again started out by setting only elements near the star surface
to evolve with FV/FD as in Sec.~\ref{stablehy}.
We are a bit more cautious than for (S) with the initial FV/FD setup, as we
know that (B) will start moving, unlike (S). Hence, not only do we set
elements that contain the star surface to FV/FD,
but also their the nearest (von Neuman) neighbors.
These result in a comparatively thicker initial band of
FV/FD elements at the surface of the star.
The star is placed at a corner of the $xy$-plane and the
$\vec{v}_\mathrm{CoM}$ is imparted diagonally along the plane to
keep the star longer in the cubical patch.
The TEI kicks in during the evolution, and we see
that only the surface of the star is being
evolved with FV/FD in the next plot
at $t = 40 M_\odot$ on the top right.
This switch to FV/FD only on the
surface happens early, near the very beginning of the simulation.
There is a small amount of matter with density $10^{-3} \rho_{0,\mathrm{C}}
(t=0)$ that has spread out, which is also evolved with FV/FD. 
We also see that the refinement spheres
follow the star center. 
In the lower left plot at time $t = 88 M_\odot$, we see that the
star has reached the corner of the central cubical region, which is the
best resolved region.
The TEI continues detecting the troubled region and
evolve this with FV/FD scheme, while maintaining DG in the rest of the mesh.
In the last (lower right) plot at $t = 132 M_\odot$,
we see that the star has crossed over from the central cube to the adjacent
cubed sphere region. Since these regions are less resolved, and since
interpolation takes place at the surface
where the central cube and the adjacent
cubed spheres meet, there is a loss of accuracy when matter
flows from the central cube to the adjacent cubed sphere patches.
Hence, we do not plan to infer any significant conclusions
from the results obtained after the point
when the majority of the star has moved out of the central cube.
Having said that, we observe that \nmesh~is able
to gracefully handle the transition of the
star into these lower resolved regions, and the simulation keeps running
smoothly, with the TEI still maintaining the FV/FD scheme
only in the narrow band near the star surface.

\begin{figure}
	\includegraphics[width=\linewidth]{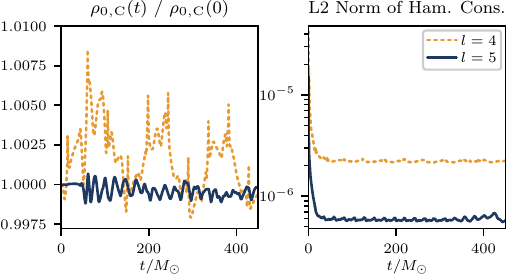}
	\caption{\label{boostRhoHam} This figure shows central density
	$\rho_{0,\mathrm{C}}$ normalized to the initial value (left) and central
	cube Hamiltonian constraint L2 norm (right) for (B) with DG+FV/FD with 
	h-refinement levels $l = 4$ and $l = 5$ for
  $v_\mathrm{CoM} = 0.1$.
}
\end{figure}

Lastly, in Fig.~\ref{boostRhoHam}
we look at the central density $\rho_{0,\mathrm{C}}$(left) and the
L2 norm of the Hamiltonian constraint in the central cube for the
$v_\mathrm{CoM} = 0.1$ case
for the duration of time for which the star remains in the central cube.
We choose $v_\mathrm{CoM} = 0.1$ so that
the star remains in the central
cube for much longer, and hence provides reliable results for
a longer duration.
We plot results for the two resolutions with
central refinement sphere h-refinment at $l = 4$ and $l = 5$.
We see both the $\rho_{0,\mathrm{C}}$ and Hamiltonian constraint
behavior improve with resolution. The results however are affected by the
fact that the $l = 4$ resolution is too low and hence the entire star is
evolved with FV/FD only, as the elements are too large and all contain
troubled regions. We leave exploration of behavior with higher
resolutions to future work.

The main takeaway from these simulations with (B) is that with
the proper setup, our TEI works efficiently and
evolves only troubled elements with FV/FD,
while using DG in all smooth regions.
The adaptive mesh refinement tracking the center of star and
the center tracking mechanism also work well. All of this is especially
true for the central cubical uniform resolution region, which is where
we plan to contain our stars in case of simulations where we want to extract
any important physics results. However, even when the star leaves this region,
\nmesh~is able to properly handle the evolution of the star as it moves into
less and less  resolved regions and crosses patch boundaries.


\section{Conclusions} 
\label{conclusions}

In this work we present a new compact FV/FD method that can be easily
combined with a DG scheme.
The aim is to construct a hybrid scheme which uses the
DG method in all regions with smooth matter fields, but which
switches to FV/FD in regions with nonsmooth matter, such as shocks
or star surfaces.

The strength of our compact FV/FD method is that it parallelizes as well as
the DG method, because we exchange information between neighboring elements
only from the surface of the element.
Through this, we minimize the amount of
data that needs to be exchanged, and we also ensure that information needs to
be exchanged only between nearest neighbors. This should allow for high
scalability when parallelizing, even while using the FV/FD method. This is a
key difference between our method and other such hybrid schemes. For
example, in~\cite{Bugner2018a} 
or~\cite{Deppe:2021ada,Deppe:2021bhi,Deppe:2024ckt}, while the
hybridization concept is similar, the FD elements still need ghost zones
consisting of points deeper than just the surface from the adjacent elements
to exchange information. Furthermore, as soon as one uses a mesh that
contains more than simple cuboid patches,
the FD method in~\cite{Bugner2018a} or~\cite{Deppe:2024ckt} requires
information exchange between all 26 neighbors, while our method requires
information exchange only between the six nearest neighbors.
In addition, the information exchange between adjacent elements simplifies
compared to a traditional FV implementation. Since we always have grid
points at the interface, we only need surface interpolation. As a result
it is much easier to implement AMR.

The price for not having additional ghost points is that near element
boundaries our compact FV/FD method is only second-order accurate.
While second-order convergence may not sound too impressive, we note that in
shocked regions we cannot expect more than first-order convergence in any
case~\cite{Godunov1959a}.
Thus, our method is adequate when used in elements with
shocks. In this way we obtain benefits of both FV/FD and DG as suggested
in~\cite{Bugner2018a,Deppe:2021ada,Deppe:2021bhi}. 

We run a battery of tests with our \nmesh~program to verify how well our new
method works in practice. We find that the compact FV/FD method indeed reaches
second-order accuracy for tests with smooth scalar fields, and that it also
performs well in a shock tube test. We then move on to tests with neutron
stars in full GRHD. As already described in~\cite{Tichy:2022hpa}, neutron
stars are more challenging due to the presence of vacuum or low-density
regions outside the star. We describe in detail the positivity limiters and
the atmosphere treatment that we have developed to deal with this
challenge. Our combination of limiters and atmosphere are successful in the
sense that they enable us to evolve neutron stars with the DG method alone.
However, the stars are then plagued by Gibbs
oscillations~\cite{Wilbraham1848a,Gibbs1898a,Gibbs1899a,Hewitt1979} that
cause the stars to pulsate. This is especially prominent when the stars are
moving. We therefore introduce a troubled-element indicator
that uses a list of trouble criteria designed to detect
elements that are not giving good results with the DG method. Using this
TEI, we build a hybrid scheme that switches between DG
and FV/FD as needed. In particular, if an element is marked as troubled and
was using DG, we redo the time step with FV/FD. In this way we try to avoid
Gibbs oscillations from ever occurring. In case a troubled element is using
FV/FD already we remain with FV/FD. On the other hand, if an element is
marked as not troubled for 11 consecutive steps and was using FV/FD, we
switch to DG, while we remain with DG in case the element was already using
it. In DG elements we use only half the number of points per direction as
in FV/FD elements, because DG is able to
provide higher accuracy results at lower cost due to its higher rate of
convergence.

Our hybrid method works well with neutron stars. In simulations
of stable TOV stars (S) where the central density should be constant,
we obtain much better results than with the pure DG method. 
If these stars are perturbed (P), they oscillate with the expected
frequency. We also show clean evolutions of boosted stars (B) and of
stars that migrate (U) from an unstable initial state
to a stable TOV star. For stable stars with and
without boost, only elements near the star surface switch to FV/FD, while
inside the stars and away from the stars no trouble is flagged so that
the more efficient DG method can be used. This means our hybrid approach
succeeds in automatically selecting the DG method in smooth regions.

We believe that this is the first time a DG method has been used to evolve
the more challenging boosted star (B) and migration (U) cases in full 3D.
In case (U), an overdense unstable star gets perturbed to
rapidly expand for a while, but later recontracts. The ensuing
strong star pulsations contain actual shocks, besides the nonsmooth
star surface. Again our hybrid method seems capable of selecting
the correct method. The bulk of the star interior and regions further from
the star still use the DG method as desired, while FV/FD is selected near
the surface and in shocked regions.

Thus, we validate that our hybrid DG+FV/FD method
is capable of handling the challenging
cases in full 3D. This gives us confidence that we have all the
necessary ingredients to evolve binary neutron star inspirals. Current work
with \nmesh~is in progress for BNS evolutions, and we hope to present
results in the near future.


\acknowledgments

It is a pleasure to thank Nils Deppe for helpful discussions.
This work was supported by NSF Grant No.
PHY-2136036 and No. PHY-2408903 and PeX-FCT (Portugal) Grant No. 2023.12549.PEX.
We also acknowledge usage of computer time on the HPC clusters KoKo
and Athene at
Florida Atlantic University, as well as on Bridges-2, Expanse and Anvil
under ACCESS allocations PHY220018 and PHY230109.
The figures in this article were created using
\texttt{matplotlib}~\cite{Hunter2007a,Caswell2020a}
and \texttt{TikZ}~\cite{Tantau2008a}.


\section*{Data Availability}

The data that support the findings of this article are
openly available~\cite{Adhikari2025b}.


\appendix

\section{Important parameter specifications}
\label{pars}

Here we state the values of the most important parameters
for the different classes of simulations for (S), (P), (U)
and (B) that we have discussed in
Sec.~\ref{results}. These are listed in Table \ref{parlist}.
By classes, we mean the different sections described above. Hence,
they are simply referred to by their section numbers as:
\S\ref{stableDG}: (S) with pure DG; \S\ref{stablehy}: (S) with DG+FV/FD,
\S\ref{perturb}: (P) with pure DG; \S\ref{fvmigpref}: (U) with pure FV/FD
and varying p-refinement; \S\ref{mighref}: (U) with pure FV/FD and
varying h-refinement; \S\ref{mighy}: (U) with DG+FV/FD; and \S\ref{boost}:
(B) with DG+FV/FD. We are using the shorthand notation ``\S''
in place of ``Sec.'' here to denote the sections.
Of course \nmesh~ has many more parameters 
that we choose not to list here.
When same parameter specification is used across multiple classes,
we group those classes together in the table.
In cases where the same specification is used for all simulations,
we simply say ``all'' in the ``classes'' column.
Some parameter choices are used for some runs, which are
not used for others entirely.
In such cases, it is likely that some other parameter is playing
an analogous role.
For example, when we set the parameter
\verb|GHG_gammas_type| to \verb|constants|, then we are using the constant
constraint damping, and we specify the values of the parameters
\verb|GHG_gamma0| $= 0.1$, \verb|GHG_gamma1| $= -1$
and \verb|GHG_gamma2| $= 1$.
However, when we set \verb|GHG_gammas_type| to \verb|3gaussians|, as in we
set the position dependent triple Gau\ss ian constraint damping functions
from Eq.~(\ref{3gaussians}), then we have to use the counterpart
parameters for these functions.
We set these values to:

\be
\begin{array}{rclrcl}
  C_{(0)} &=& 0.01, & & & \\
  A^{(0)}_{(1)} &=& 0.06277857994, &w^{(0)}_{(1)} &=& 7.884855, \\
  A^{(0)}_{(2)} &=& 0, &w^{(0)}_{(2)} &=& 7.884855,\\
	A^{(0)}_{(3)} &=& 0.06277857994, &w^{(0)}_{(3)} &=& 51.60996,
  \label{Par3Gf}
\end{array}
\ee

\be
\begin{array}{rclrcl}
  C_{(1)} &=& -0.999, & & & \\
  A^{(1)}_{(1)} &=& 0, &w^{(1)}_{(1)} &=& 1, \\
	A^{(1)}_{(2)} &=& 0, &w^{(1)}_{(2)} &=& 1,\\
	A^{(1)}_{(3)} &=& 0.06277857994, &w^{(1)}_{(3)} &=& 318.534,\\
\end{array}
\ee

\be
\begin{array}{rclrcl}
  C_{(2)} &=& 0.01, & & & \\
	A^{(2)}_{(1)} &=& 0.94167869922, &w^{(2)}_{(1)} &=& 7.884855, \\
	A^{(2)}_{(2)} &=& 0, &w^{(2)}_{(2)} &=& 7.884855,\\
	A^{(2)}_{(3)} &=& 0.19182343873, &w^{(2)}_{(3)} &=& 51.60996.
  \label{Par3Gl}
\end{array}
\ee

$A^{(0)}_{(2)}$ and $A^{(2)}_{(2)}$ are set to $0$ in our case, as unlike
in~\cite{Deppe:2024ckt}, we have only one star in our simulations.
We name the parameters corresponding to the constants $C_{(0)}$,
$A^{(0)}_{(1)}$ and $w^{(0)}_{(1)}$ as \verb|GHG_3gauss_gamma0_C|,
\verb|GHG_3gauss_gamma0_A1| and \verb|GHG_3gauss_gamma0_w1| respectively,
and the name of the rest of the parameters follow suite. Owing to the large
number of constants, we do not repeat these position-dependent constarint
damping parameters in the table. All the run classes that set
\verb|GHG_gammas_type| = \verb|3gaussians| use these values for the constants.

\begin{table*}
	\caption{\label{parlist} List of selected crucial parameters and their
  values for the different classes of simulations. The first column states
  the name of the par as it is registered in \nmesh~along with a brief
  description. The second and third column provides the specification of the
  par and the simulation classes in which that specification is used.
  The specifications of the parameters stated in this table are kept same
  for all the simulations in a given class.
	} 
	\centering
	\begin{tabular}{p{0.4\linewidth} p{0.25\linewidth} p{0.31\linewidth}}
	\hline
  \hline
	Name and description & Specification & Simulation class
	\\ \hline
	\hline
  \verb|EoS_PwP_kappa|: 
  Value of $\kappa$ in Eq~.(\ref{eos}).
      & \verb|100|
          & all
  \\ \hline
  \verb|EoS_PwP_n|:  Value of $n$ in Eq.~(\ref{eos}).
      & \verb|1|
          & all
  \\ \hline
  \verb|EoS_type|:  Type of matter EoS.
      & \verb|ideal|
          & all
  \\ \hline
  \verb|GHG_fd_dissfac|:  Value of $\sigma$ in Eq.~(\ref{KO-diss}) \newline
            in FV/FD elements for GH system.
      & \verb|0.1|
          & \S\ref{stablehy}, \S\ref{fvmigpref}, 
            \S\ref{mighref}, \S\ref{mighy},  \S\ref{boost}
  \\ \hline
  \verb|GHG_fd_dissorder|: Dissipation order in \newline
        Eq.~(\ref{KO-diss}) in the interior of FV/FD elements.
      & \verb|4|
          & \S\ref{stablehy}, \S\ref{fvmigpref}, 
            \S\ref{mighref}, \S\ref{mighy},  \S\ref{boost}
  \\ \hline
  \verb|GHG_fd_dissorder_min|: Minimum \newline dissipation order in 
        Eq.~(\ref{KO-diss}) allowed 
      & \verb|4|
          & \S\ref{stablehy}, \S\ref{mighref},
             \S\ref{mighy},  \S\ref{boost} \\ \cline{2-3}
   near boundaries in FV/FD elements.
      &  \verb|2|
          & \S\ref{fvmigpref}
  \\ \hline  
  \verb|GHG_gammas_type|: Type of
      & \verb|constants|
          & \S\ref{stableDG}, \S\ref{stablehy}, \S\ref{perturb} \\ \cline{2-3}
  constraint damping functions: \newline
  constants or position dependent Gau\ss ians.
      &  \verb|3gaussians|
          & \S\ref{fvmigpref}, \S\ref{mighref}, 
          \S\ref{mighy},  \S\ref{boost}
  \\ \hline  
  \verb|GHG_gauge|: Gauge choice: 
      & \verb|freeze|
          & \S\ref{stableDG}, \S\ref{stablehy}, \S\ref{perturb} \\ \cline{2-3}
  set Eq.~(\ref{freeze})(\verb|freeze|) or set $H_a = 0$ \newline
  by setting amplitude of \verb|damped_wave| to $0$.
      &  \verb|damped_wave|
          & \S\ref{fvmigpref}, \S\ref{mighref}, 
          \S\ref{mighy},  \S\ref{boost}
  \\ \hline 
  \verb|GHG_mu0|: Amplitude for  \verb|damped_wave|. \newline
  Setting this to \verb|0| sets Harmonic gauge.
      & \verb|0|
          & \S\ref{fvmigpref}, \S\ref{mighref}, 
      \S\ref{mighy},  \S\ref{boost}
  \\ \hline 
  \verb|GRHD_Persson_alpha|: Value of $\alpha_N^{\mathrm{evo}}$ \newline
  from Eq.~(\ref{persson}) in DG elements.
      & \verb|8|
          & \S\ref{stablehy},\S\ref{mighy},  \S\ref{boost}
  \\ \hline 
  \verb|GRHD_Persson_alpha_fv|: Value of $\alpha_N^{\mathrm{evo}}$ \newline
  from Eq.~(\ref{persson}) in FV/FD elements.
      & \verb|9|
          & \S\ref{stablehy},\S\ref{mighy},  \S\ref{boost}
  \\ \hline 
  \verb|GRHD_Persson_f_unfilt|: Value of \newline $f_\mathrm{unfiltered}$
  related to Eqs.~(\ref{ncsurf},\ref{ncvol}.)
      & \verb|0.9|
          & \S\ref{stablehy},\S\ref{mighy},  \S\ref{boost}
  \\ \hline 
  \verb|GRHD_S_filter_alp|, \verb|GRHD_filter_alp|, \newline
  \verb|evolve_filter_alp|: 
  Value of $\alpha_f$  for exponential filters for all filtered variables
  in DG elements.
      & \verb|36|
          & \S\ref{stableDG}, \S\ref{perturb},
          \S\ref{stablehy}, \S\ref{mighy}, \S\ref{boost} 
  \\ \hline
  \verb|GRHD_S_filter_s|, \verb|GRHD_filter_s|: Value 
      & \verb|32|
          & \S\ref{stableDG}, \S\ref{perturb} \\ \cline{2-3}
  of $s$ for exponential filter applied to \newline
  $D$, $\tau$ and $S_i$ in DG elements.
      & \verb|40|
          & \S\ref{stablehy}, \S\ref{mighy}, \S\ref{boost} 
  \\ \hline 
  \verb|GRHD_S_limfac|: Scaling factor $f_S^\mathrm{lim}$ for \newline
  pointwise limiting of $S_i$ in Sec.~\ref{sec:lim}.
      & \verb|0.99|
          & \S\ref{fvmigpref}, \S\ref{stablehy},
          \S\ref{mighref}, 
          \S\ref{mighy}, \S\ref{boost} 
  \\ \hline 
  \verb|GRHD_Tau_atmolimfac|: Used with \newline \verb|limit_negTau| as value
  $f_{\tau,\mathrm{atmo}}^\mathrm{lim}$ in atmosphere \newline
  such that $\tau$ is
  limited to be $\tau >= - f_{\tau,\mathrm{atmo}}^\mathrm{lim} \cdot D $.
      & \verb|0|
          & \S\ref{fvmigpref}
  \\ \hline 
  \verb|GRHD_Tau_limfac|: Used with  \verb|limit_negTau| \newline
  as value
  $f_{\tau}^\mathrm{lim}$ outside atmosphere such \newline that $\tau$ is
  limited to be $\tau >= - f_{\tau}^\mathrm{lim} \cdot D $.
      & \verb|0.1|
          & \S\ref{fvmigpref}
  \\ \hline 
  \verb|GRHD_atmo_Wvmax|: Scaling factor \newline $(Wv)_\mathrm{atmo,max}$ for 
  $Wv$ from Sec.~\ref{atmo}.
      & \verb|0.01|
          & all
  \\ \hline 
  \verb|GRHD_atmo_cut|: Factor $f_\mathrm{atmo,cut}$ 
      & $\mathtt{1.01 \times 10^{-12}}$
          & \S\ref{stableDG}, \S\ref{perturb},  \\ \cline{2-3}
  from Sec.~\ref{atmo}.
      & $\mathtt{10^{-9}}$
          & \S\ref{fvmigpref}, \S\ref{stablehy}, 
          \S\ref{mighref}, 
          \S\ref{mighy}, \S\ref{boost} 
  \\ \hline 
  \verb|GRHD_atmo_epslmax|: Factor $\epsilon_\mathrm{atmo,max}$ \newline
  from Sec.~\ref{atmo}.
      & \verb|0.1|
          & all
  \\ \hline 
  \verb|GRHD_atmo_level|: Factor $f_\mathrm{atmo,level}$ 
      & $\mathtt{10^{-12}}$
          & \S\ref{stableDG}, \S\ref{perturb},  \\ \cline{2-3}
  from Sec.~\ref{atmo}.
      & $\mathtt{10^{-11}}$
          & \S\ref{fvmigpref}, \S\ref{stablehy}, 
          \S\ref{mighref}, 
          \S\ref{mighy}, \S\ref{boost}
  \\ \hline
  \verb|GRHD_atmo_max_fac|: Factor $f_\mathrm{atmo,max}$ 
      & \verb|10|
          & \S\ref{stableDG}, \S\ref{perturb},  \\ \cline{2-3}
  from Sec.~\ref{atmo}.
      & \verb|1|
          & \S\ref{fvmigpref}, \S\ref{stablehy}, 
          \S\ref{mighref}, 
          \S\ref{mighy}, \S\ref{boost}
  \\ \hline
  \end{tabular}
\end{table*}
\begin{table*}
  \begin{tabular}{p{0.4\linewidth} p{0.25\linewidth} p{0.31\linewidth}}
  \hline
  \verb|GRHD_cons2prim_Wvmax_fac|: Widening \newline factor for the $Wv$
  root finding bracket.
      & \verb|1.1|
          & \S\ref{fvmigpref}, \S\ref{stablehy}, \S\ref{mighref}, 
      \S\ref{mighy},  \S\ref{boost}
  \\ \hline 
  \verb|GRHD_cons2prim_epslfloor|: Minimum \newline $\epsilon$ value allowed
  when converting
      & \verb|0|
          &  \S\ref{stablehy}, \S\ref{mighref},  
      \S\ref{mighy},  \S\ref{boost} \\ \cline{2-3} 
  from $(D, \tau, S_i)$ to $(\rho, \epsilon, P, Wvi^i)$.
      & \verb|-0.1|
          & \S\ref{fvmigpref}
  \\ \hline 
  \verb|GRHD_cs2_limited|: Upper limit for \newline square of sound speed.
      & \verb|1|
          & \S\ref{stablehy},  \S\ref{mighref}, 
      \S\ref{mighy},  \S\ref{boost}
  \\ \hline 
  \verb|GRHD_limiter|: Positivity limiter  
      & \verb|limit_S| \verb|limit_Tau| \verb|limit_D|
          & \S\ref{stableDG}, \S\ref{perturb} \\ \cline{2-3} 
  choices (Sec.~\ref{sec:lim}).
          & \verb|try_DTS_limit| \verb|limit_S|
        \newline \verb|limit_negTau| \verb|limit_D|
          &  \S\ref{fvmigpref} \\\cline{2-3} 
      & \verb|try_DTS_limit| \verb|limit_S|
        \newline \verb|limit_Tau| \verb|limit_D|
          & \S\ref{stablehy}, \S\ref{mighref},
          \S\ref{mighy}, \S\ref{boost}
  \\ \hline
  \verb|GRHD_limiter_atmo_level| : Atmosphere level \newline
  factor $f^\mathrm{lim}_\mathrm{atmo,level}$ in
  Sec.~\ref{sec:lim}, such that \newline
  $\rho^\mathrm{lim}_\mathrm{atmo,level}
  = f^\mathrm{lim}_\mathrm{atmo,level} \cdot \rho_{0,\mathrm{max}}(t=0)$.
      & $\mathtt{10^{-12}}$
          & all 
  \\ \hline
  \verb|GRHD_numflux|:
  Numerical flux for hydro.
      & \verb|LLF|
          & all
  \\ \hline
  \verb|GRHD_rec_epslmax_fac|: Upper $\epsilon$ cutoff factor \newline
  for region with
  $w =1$ in Eq.~(\ref{fv_divf_extrap}).
      & \verb|1000|
          & \S\ref{stablehy},  \S\ref{mighref}, 
      \S\ref{mighy},  \S\ref{boost}
  \\ \hline 
  \verb|GRHD_rec_epslmin_fac|: Lower $\epsilon$ cutoff factor \newline
  for region with
  $w =1$ in Eq.~(\ref{fv_divf_extrap}).
      & $\mathtt{10^{-9}}$
          & \S\ref{stablehy},  \S\ref{mighref}, 
      \S\ref{mighy},  \S\ref{boost}
  \\ \hline 
  \verb|GRHD_rec_epslmin_fac|: lower $\rho_0$ cutoff factor \newline
  for region with
  $w =1$ in Eq.~(\ref{fv_divf_extrap}).
      & $\mathtt{10^{-9}}$
          & \S\ref{stablehy},  \S\ref{mighref}, 
      \S\ref{mighy},  \S\ref{boost}
  \\ \hline 
  \verb|GRHD_trouble_n_fv|: Points in each direction \newline
  in FV/FD elements compared
  to DG: FV/FD \newline
  has twice that in DG in each direction if \verb|2n_dg|.
      & \verb|2n_dg|
          & \S\ref{stablehy}, \S\ref{mighy},  \S\ref{boost}
  \\ \hline 
  \verb|GRHD_troublescore_cut|: Factor $f_\mathrm{cut,low}$ related to
  Eq.~(\ref{TroubleD}).
      & $\mathtt{10^{-4}}$
          & \S\ref{stablehy}, \S\ref{mighy},  \S\ref{boost}
  \\ \hline 
  \verb|GRHD_troublescore_cut_fac|: Factor $f_\mathrm{cut,hi}$ related to
  Eq.~(\ref{TroubleD}).
      & \verb|10|
          & \S\ref{stablehy}, \S\ref{mighy},  \S\ref{boost}
  \\ \hline 
  \verb|TOVstar_boost_Gammav|: $Wv^i$ components of star's CoM in intial data.
      & \verb|0 0 0|
          & \S\ref{stableDG}, \S\ref{perturb},  
          \S\ref{fvmigpref},  \S\ref{stablehy},   \newline
          S\ref{mighref}, \S\ref{mighy} \\ \cline{2-3}
      & \verb|0.0710669| \verb|0.0710669| \verb|0.0|, \newline
        \verb|0.4082482| \verb|0.4082482| \verb|0.0|
          &  \S\ref{boost}
  \\ \hline
  \verb|TOVstar_central_quantity|: Field whose \newline
  central value sets initial data.
      & \verb|rho0_c|
          & all
  \\ \hline 
  \verb|TOVstar_central_value|: Initial value of
      & \verb|0.00128|
          & \S\ref{stableDG}, \S\ref{perturb},  
            \S\ref{stablehy},  \S\ref{boost}   \\ \cline{2-3}
  central quantity $\rho_{0,\mathrm{C}}$.
      & \verb|0.0079934|
          &  \S\ref{fvmigpref},  \S\ref{mighref},  \S\ref{mighy}
  \\ \hline
  \verb|TOVstar_refine_AboveCoeffFallOff|: \newline
  Exponential coefficient fall off
  cutoff used to \newline
  set surface of star to FV/FD in intial data.
      & \verb|0.8|
          & \S\ref{stablehy}, \S\ref{boost}
  \\ \hline
  \verb|TOVstar_fv_n0|, \verb|TOVstar_fv_n1|, \newline
  \verb|TOVstar_fv_n2|:
  Initial number of points \newline in FV/FD elements in TOV star setup.
      &  \verb|16|
          & \S\ref{stablehy}, \S\ref{boost}
  \\ \hline
  \verb|TOVstar_perturb|: Which quantity to
      &  \verb|P|
					& \S\ref{perturb}  \\ \cline{2-3}
	perturb.
			& \verb|rho0 only|
					& \S\ref{mighy}
  \\ \hline
  \verb|TOVstar_perturb_l|: Value for $\ell$ for
      &  \verb|2|
					& \S\ref{perturb}  \\ \cline{2-3}
	spherical harmonic $Y^m_\ell(\theta,\varphi)$ in\newline 
	perturbation function (see Eq.~(\ref{pertP})). 
			& \verb|0|
					& \S\ref{mighy}
  \\ \hline
  \verb|TOVstar_perturb_m|: Value for $m$ for
      &  \verb|0|
					& \S\ref{perturb}  \\ \cline{2-3}
	spherical harmonic $Y^m_\ell(\theta,\varphi)$ in\newline 
	perturbation function (see Eq.~(\ref{pertP})).
			& \verb|0|
					& \S\ref{mighy}
  \\ \hline
\end{tabular}
\end{table*}
\begin{table*}
  \begin{tabular}{p{0.4\linewidth} p{0.25\linewidth} p{0.31\linewidth}}
  \hline
  \verb|TOVstar_perturb_n|: Value for $n$ in
      &  \verb|1|
					& \S\ref{perturb}  \\ \cline{2-3}
	in $ \sin [ \{ (n+1)\pi r / 2 r_\mathrm{surf} \} + R ]$ in
	\newline perturbation function (see Eq.~(\ref{pertP})).
			& \verb|0|
					& \S\ref{mighy}
  \\ \hline
  \verb|TOVstar_perturb_rphase|: Value for $R$ in
      &  \verb|0|
					& \S\ref{perturb}  \\ \cline{2-3}
	in $ \sin [ \{ (n+1)\pi r / 2 r_\mathrm{surf} \} + R ]$ in
	\newline perturbation function (see Eq.~(\ref{pertP})).
			& $\mathtt{\pi/2}$
					& \S\ref{mighy}
  \\ \hline
  \verb|TOVstar_perturb_lambda|: Value for $\lambda$ in
      &  \verb|0.05|
					& \S\ref{perturb}  \\ \cline{2-3}
	in perturbation function (see Eq.~(\ref{pertP})).
			& \verb|-0.01|
					& \S\ref{mighy}
  \\ \hline
	\verb|amr_BoxMesh_dout|:  Dimension of the
      & \verb|16|
          & \S\ref{stableDG},  \S\ref{perturb} \\ \cline{2-3}
  central cube.
      & \verb|8|
          & \S\ref{fvmigpref}, \S\ref{mighref},
             \S\ref{mighy}  \\ \cline{2-3}
      & \verb|26|
          & \S\ref{stablehy}, \S\ref{boost}
  \\ \hline
  \verb|amr_BoxMesh_npatches|: 
  Number of \newline cubical patches in the mesh.
      & \verb|1|
          & all
  \\ \hline
  \verb|amr_CubedSphere_ndomains|:
  Number of  \newline  cubed sphere patches in the mesh.
      & \verb|6|
          & all
  \\ \hline
  \verb|amr_CubedSphere_r0fac|:
  Factor by \newline which radius
  of cubed sphere is larger 
  than \newline central cube dimension.
      & \verb|4|
          & all
  \\ \hline
  \verb|amr_Shell_nr|: 
    Number of spherical shells \newline in radial direction.
      & \verb|1|
          & all 
  \\ \hline
  \verb|amr_Shell_rout|:  
      & \verb|320|
          & \S\ref{stableDG}, 
          \S\ref{perturb},  \S\ref{fvmigpref} \\  \cline{2-3}
	Radius of outer boundary.
      & \verb|200|
          & \S\ref{mighref}, \S\ref{mighy} \\  \cline{2-3} 
      & \verb|208|
          &  \S\ref{stablehy}, \S\ref{boost} 
  \\ \hline
  \verb|amr_mesh_type|: Types of patches we use.
      & \verb|BoxMesh| \verb|CubedSpheres| \verb|Shell|
          & all
  \\ \hline
  \verb|amr_hrefine_p| : 
  Number of Patch to be \newline h-refined  one time more than rest.
      & \verb|0|
          & \S\ref{stableDG}, \S\ref{perturb}
  \\ \hline
  \verb|amr_luni|: Uniform minimum
      & \verb|3|,\verb|4|,\verb|5|
          & \S\ref{stableDG},  \S\ref{perturb} \\ \cline{2-3} 
  h-refinement  applied to  mesh.
      & \verb|3|
          &  \S\ref{stablehy}, \S\ref{fvmigpref} 
  \\ \hline
  \verb|amr_n0|, \verb|amr_n1| \verb|amr_n2|: initial number of 
      & \verb|5|
          &  \S\ref{stableDG}, \S\ref{perturb}, \\ \cline{2-3}
  points in general in elements.
      & \verb|16|
          & \S\ref{stablehy}, \S\ref{mighy}, \S\ref{boost}
  \\ \hline
  \verb|basis_filter_fv|: Whether to filter FV/FD \newline element fields.
      & \verb|no| 
          & \S\ref{stablehy},  \S\ref{fvmigpref} \newline
          \S\ref{mighref}, \S\ref{mighy},  \S\ref{boost}
  \\ \hline 
  \verb|center1_amr_lmax|: Maximum
			& \verb|5|
					&  \S\ref{stablehy}  \\  \cline{2-3}
	h-refinement in central sphere when using
      & \verb|3|,   \verb|4|,  \verb|5|
          & \S\ref{mighref} \\  \cline{2-3}
	concentric spherical refinement regions.
			& \verb|3|, \verb|4|
          &\S\ref{mighy}, \\  \cline{2-3}
      & \verb|4|, \verb|5|
          & \S\ref{boost}
  \\ \hline 
  \verb|center1_amr_radius|: Radius of 
      & \verb|30.5|  
          & \S\ref{mighref}, \S\ref{mighy}, \\  \cline{2-3}
  innermost highest h-refined sphere.
      & \verb|15|
          & \S\ref{stablehy}, \S\ref{boost}
  \\ \hline 
  \verb|center1_track|: Track extremum to track \newline star center. 
      & \verb|min|  
          & \S\ref{boost}
  \\ \hline 
  \verb|center1_track_var|: Variable whose \newline extremum is tracked. 
      & \verb|ADM_alpha|  
          & \S\ref{boost}
  \\ \hline 
  \verb|dt|: Time step.
      & \verb|0.1|, \verb|0.05|, \verb|0.025|
          & \S\ref{stableDG},  \S\ref{perturb} 
  \\ \hline 
  \verb|dtfac|: Value of $1/v$ in Eq.~(\ref{cfl}) for DG elements.
      & \verb|0.25|
          &   \S\ref{stablehy},  \S\ref{fvmigpref} 
          \S\ref{mighref}, \S\ref{mighy},  \S\ref{boost}
  \\ \hline 
  \verb|evolve_filter_s|: Value of $s$ for exponential
      & \verb|40|
          & \S\ref{stableDG}, \S\ref{perturb}, \S\ref{fvmigpref} \\ \cline{2-3}
   filter applied to $g_{ab}$, $\Pi_{ab}$ and $\Phi_{iab}$ in
  DG elements.
      & \verb|64|
          & \S\ref{stablehy}, S\ref{mighref}, 
          \S\ref{mighy}, \S\ref{boost} 
  \\ \hline 
  \verb|evolve_method|: RK evolution scheme: \newline
  stability preserving 3rd order
  if \verb|sspRK3|.
      & \verb|sspRK3|  
          & all
  \\ \hline 
  \verb|fd_stencilsize|: Total points in FD stencils.
      & \verb|3|
          & \S\ref{stablehy}, \S\ref{fvmigpref}, 
            \S\ref{mighref}, \S\ref{mighy},  \S\ref{boost}
  \\ \hline
  \verb|fv_divf_extrap|: Extrapolation usage in 
      & \verb|no|
          &  \S\ref{fvmigpref} \\ \cline{2-3} 
  eqn(\ref{fv_divf_extrap}): \verb|no| sets $w=1$; $w=4/3$ otherwise.
          & \verb|dnfn_extrap1| 
          & \S\ref{stablehy}, \S\ref{mighref},
          \S\ref{mighy}, \S\ref{boost}
  \\ \hline
\end{tabular}
\end{table*}
\begin{table*}
  \begin{tabular}{p{0.4\linewidth} p{0.25\linewidth} p{0.31\linewidth}}
  \hline
  \verb|fv_rec|: Reconstruction in FV elements: \newline
  ``WENOZ\ts{2}'' described in
  \S\ref{modFV}.
      & \verb|WENOmZ_2|
          & \S\ref{stablehy}, \S\ref{fvmigpref},
            \S\ref{mighref}, \S\ref{mighy},  \S\ref{boost}
  \\ \hline
  \verb|fv_surface_interp|: Interpolation at 
      & \verb|linear|
          &  \S\ref{fvmigpref} \\ \cline{2-3} 
  boundary between neighboring \newline
  FV elements to exhange data.
      & \verb|parabolic| 
          & \S\ref{stablehy}, \S\ref{mighref},
           \S\ref{mighy}, \S\ref{boost}
  \\ \hline
  \verb|uniform_dtfac|: Value of $1/v$ \newline
  in Eq.~(\ref{cfl}) for FV/FD elements.
      & \verb|0.125|
          & \S\ref{stablehy}, \S\ref{fvmigpref},
            \S\ref{mighref}, \S\ref{mighy},  \S\ref{boost}
  \\ \hline
	\end{tabular}
\end{table*}


\bibliography{references}

\end{document}